\documentclass[twocolumn,english,pre,floatfix,citeautoscript,nofootinbib]{revtex4}
\usepackage{amsbsy}
\usepackage{latexsym,epsfig,graphicx}
\usepackage{dcolumn}
\usepackage{subfigure}
\usepackage{comment}
\usepackage{color}
\usepackage[colorlinks,urlcolor=blue,citecolor=blue]{hyperref}
\usepackage{amstext}
\usepackage{amssymb}
\usepackage{setspace}
\usepackage{amsmath}
\usepackage{makeidx}
\usepackage{bm}
\usepackage{ulem}
\usepackage{multirow}
\usepackage{mathrsfs}

\usepackage{array}

\begin{document}

\title{Phase Characters of Optical Dark Solitons with the Third-order Dispersion and Delayed Nonlinear Response}
\author{Yan-Hong Qin$^{1}$}
\author{Xiaoman Zhang$^{1}$}
\author{Liming Ling$^{1}$}\email{linglm@scut.edu.cn}
\author{Li-Chen Zhao$^{2,3,4}$}\email{zhaolichen3@nwu.edu.cn}

\address{$^{1}$School of Mathematics, South China University of Technology, Guangzhou 510640, China}
\address{$^{2}$School of Physics, Northwest University, Xi'an 710127, China}
\address{$^{3}$Shaanxi Key Laboratory for Theoretical Physics Frontiers, Xi'an 710127, China}
\address{$^{4}$Peng Huanwu Center for Fundamental Theory, Xian 710127, China}
\begin{abstract}
Dark soliton is usually seen as one of the simplest topological solitons, due to the phase jump across its density dip. We investigate the phase jump properties of dark solitons in a single mode optical fiber with the third-order dispersion and delayed nonlinear response, based on exact analytical solutions of Hirota equation. Our analysis indicates that a single-valley dark soliton (SVDS) can admit two distinct phase jumps at the same velocity, in sharp contrast to the dark soliton with only the second-order dispersion and self-phase modulation, which admits a one-to-one match between the velocity and phase jump. We further uncover the different topological vector potentials underlying the distinct phase jumps. The relations between phase jump and velocity of the SVDS can explain the generation of the previously reported double-valley dark soliton (DVDS). The detailed analysis on the two phase jumps characters of the DVDS with one identical velocity enables us to obtain U-shaped type or double-step type phase distribution. We further explore collision properties of the DVDSs by analyzing their topological phase, which can be considered as the generalization of topological phase (Phys. Rev. E 103, L040204). Strikingly, the inelastic collision can lead to the conversion between the two types of phase distributions for DVDS. The results reveal that inelastic or elastic collision can be judged by analyzing the magnetic monopole fields.

\end{abstract}
\date{\today}

\maketitle

\section{Introduction}

Nonlinear optical fibers are known to be the most convenient quasi-one-dimensional system to investigate dynamical properties of solitons,  owing to the simplicity and precise control on the experimental parameters \cite{nop,Kivshar,optics1,optics2,optics3,optics4,appl1,optics5,BS2}. In the picosecond regime, the propagation of optical pulses is modeled by the one-dimensional nonlinear Schr\"{o}dinger equation (NLSE) \cite{DS1,BS1}, which accounts for the second-order dispersion and self-phase modulation. This generic model with self-defocusing (self-focusing) nonlinearity admits dark (bright) soliton \cite{BS1,BS2,DS1,optics3}. In contrast to the bright soliton, dark soliton uniquely displays one phase jump across its intensity dip \cite{nop,Kivshar}. Because of this, dark soliton is usually seen as one of the simplest topological solitons \cite{top1,top2,top3,top4}. Recently, the intriguing topological phases of the SVDS in scalar NLSE and multivalley dark soliton in vector NLSE have been revealed \cite{zhaods,Qinds}, via extending the complex coordinate space to explore the density zeros of dark solitons. The phase variations of dark solitons are usually velocity dependent. More specifically, there is a one-to-one match between the phase jump and velocity of dark solitons in NLSE \cite{nop,Kivshar,zhaods,Qinds}.

For ultrashort pulses (in the subpicosecond or femtosecond regime), the higher-order effects such as higher-order dispersion and delayed nonlinear response become important \cite{highorder1,highorder2}. Considering these effects, the optical solitons follow from the higher-order NLSE rather than the standard NLSE \cite{HE1,HE2,HE3,highorder1,highorder2,solution6}, such as the well-known Hirota equation (HE) \cite{HE1}. Recent studies have demonstrated that the single mode self-defocusing HE admits not only the SVDS but also the DVDS \cite{zhangds}, while the latter cannot be obtained in standard NLSE. The unusual inelastic collision of DVDS has also been revealed. However, the topological phases of these dark solitons in HE are far from fully understood. These motivate us to explore the phase characters of optical dark solitons with high-order effects.

In the present work, we focus on the phase properties of SVDS and DVDS in a single mode optical fiber with the third-order dispersion and delayed nonlinear response, by the aid of exact dark soliton solutions of HE. We find that these dark solitons can admit two different phase jumps at the same velocity under the proper conditions, which is in sharp contrast with the one-to-one correspondence between the phase jump and velocity of dark solitons in NLSE. The different topological vector potentials underlying the distinct phase jumps have been revealed. For SVDS, we present the existence diagrams for the phase jump and velocity meeting such relation when the background wavenumber and the higher-order nonlinearity are considered. Importantly, such relation between the phase jump and velocity of the SVDS can be used to understand the formation mechanism of DVDS reported previously. Meanwhile, it leads to double-step type or U-shaped type phase distribution of a DVDS with the same velocity. Furthermore, we discuss collision properties of DVDSs from phase standpoint, via combining the topological vector theory with the developed asymptotic analysis technique. Interestingly, the collision can change the type of phase distribution of DVDS, due to the variations of magnetic monopole fields. The phase diagrams of the phase characters for the DVDS before and after collision have been presented. The inelastic collision must result in the changes of magnetic monopole fields in the complex space. For the elastic collision, both the intensity profiles and topological phases remain unchanged. These phase properties provide an alternative approach to understand the physical mechanism and the collision properties of dark solitons.

Our paper is organized as follows. In Sec. \uppercase\expandafter{\romannumeral2}, we present the physical model and study phase characters of the SVDS. The existence phase diagrams for the two cases of relations between the phase jump and velocity have been presented. In Sec. \uppercase\expandafter{\romannumeral3}, two types of phase distributions for a DVDS with the same velocity have been studied in detail. In Sec. \uppercase\expandafter{\romannumeral4}, we discuss the collision properties of DVDSs based on the topological theory and the asymptotic analysis technique.  Finally, the summary is given in Sec. \uppercase\expandafter{\romannumeral5}.

\section{The physical model and phase characters of the single-valley dark soliton}\label{sec2}

The single mode optical fibers with the third-order dispersion and delayed nonlinear response can be governed by the well-known HE \cite{HE1,HE2,HE3}. In dimensionless form it is given by
\begin{equation}
\mathrm{i}\frac{\partial q}{\partial t}+\frac{1}{2}\frac{\partial^2 q}{\partial x^2}+\sigma \left |q \right|^2 q-\mathrm{i}\beta\left(\frac{\partial^3 q}{\partial x^3}+\sigma 6\left |q \right|^2 \frac{\partial q}{\partial x}\right)=0,
\label{eq:Hirota}
\end{equation}
where $t$, $x$ are time evolution and space distribution coordinates, and $q$ is the envelope of the
wave field. Note, however, that in the context of optical fibers the roles of $t$ and $x$ are reversed.  This model can be used to describe the  propagations of ultrashort light pulses, such as subpicosecond or femtosecond pulses. The last two terms in the Eq.~\eqref{eq:Hirota} that enter with a real coefficient $\beta$ are responsible for the third-order dispersion and delayed nonlinear response, respectively.  Based on the integrability of HE, many kinds of exact localized waves solutions have been obtained by various methods. There are dark solitons in defocusing case ($\sigma=-1$) \cite{solution6,zhangds,Lou,solution4,Hoseini}, and  bright solitons, rogue waves, rational solitons and breathers in focusing case ($\sigma=1$) \cite{solution1,solution2,solution3,solution5,solution7,Chen}.

\begin{table}[htp]
\centering
\renewcommand\arraystretch{1.5}
\setlength{\tabcolsep}{2mm}
\caption{The velocity ranges of SVDS. Here, the expressions
$v(z_1=\pi)=3(a^2+2ac+2c^2)\beta+c$, $v(z_1=0)=3(a^2-2ac+2c^2)\beta-c$, $v(z_1=z_e)=(3a^2+2c^2)\beta-\frac{(1+6a\beta)^2}{16\beta}$, and $z_e=\arccos(\frac{\rho}{2})$ with $\rho=(6a\beta+1)/(4c\beta)$. }
\footnotesize
\label{v1range}
\begin{tabular}{c|c|c}
 \hline
  $\beta$ & $a$ & $v$ \\
  \hline
  \multirow{4}{*}{$\beta>0$} & $a<-\frac{8c\beta+1}{6\beta}$ & {$\big(v(z_1=\pi),v(z_1=0)\big)$} \\
  \cline{2-3}
  &$a>\frac{8c\beta-1}{6\beta}$ & {$\big(v(z_1=0),v(z_1=\pi)\big)$} \\
  \cline{2-3}
  &$-\frac{1}{6\beta}\leq a<\frac{8c\beta-1}{6\beta}$ & {$\big(v(z_1=z_e),v(z_1=\pi)\big)$} \\
  \cline{2-3}
  &$-\frac{8c\beta+1}{6\beta}< a\leq-\frac{1}{6\beta}$ & {$\big(v(z_1=z_e),v(z_1=0)\big)$} \\
  \hline
  \multirow{4}{*}{$\beta<0$} & $a<-\frac{8c\beta+1}{6\beta}$ & {$\big(v(z_1=0),v(z_1=\pi)\big)$} \\
  \cline{2-3}
  &$a>\frac{8c\beta-1}{6\beta}$ & {$\big(v(z_1=\pi),v(z_1=0)\big)$} \\
  \cline{2-3}
  &$-\frac{1}{6\beta}\leq a<\frac{8c\beta-1}{6\beta}$ & {$\big(v(z_1=\pi),v(z_1=z_e)\big)$} \\
  \cline{2-3}
  &$-\frac{8c\beta+1}{6\beta}< a\leq-\frac{1}{6\beta}$ & {$\big(v(z_1=0),v(z_1=z_e)\big)$} \\
  \hline
\end{tabular}
\end{table}

In the present paper, we will investigate the phase characters of SVDS and DVDS in defocusing case, based on the exact $n$-dark soliton solutions $q^{[n]}$ (see Eq.~\eqref{eq:q[n]}) of the Eq.~\eqref{eq:Hirota}.
The detailed derivation processes have been given, by applying the $n$-fold Darboux transformation (DT) \cite{lingDT} on the plane wave background $q^{[0]}=c\rm{e}^{\rm{i} \theta}$ with $\theta=a x-\big[\beta(a^2+6c^2)a+1/2a^2+c^2\big]t$. The spectral parameters are set as $\lambda_i=-\frac{a}{2}+c\cos{z_i}$ with $z_i\in(0,\pi) (i=1,\ldots,n)$ to simplify the solitons solution.

\subsection{The velocity ranges}

We start with one SVDS solution of the Eq.~\eqref{eq:Hirota}, with choosing $n=1$ of Eq.~\eqref{eq:q[n]},
\begin{eqnarray}
\label{onesvds}
q^{[1]}=c[\tilde{v}_1+\mathrm{i}\tilde{w}_1\tanh[\tilde{w}_1(x-v_1t+\gamma_1)]\mathrm{e}^{\mathrm{i}\theta},
\end{eqnarray}
where $c$ and $a$ are the amplitude and wavenumber of plane wave background, respectively. We set $c=1$ thereinafter.  $\gamma_1$ is the position of SVDS in distribution direction. $\tilde{w}_1=\sin(z_1)$ is the width-dependent parameter. $v_1$ is the velocity of soliton center, calculated as
$v_1=\beta(4c^2\tilde{v}_1^2-6ac\tilde{v}_1+3a^2+2c^2)-c\tilde{v}_1+a$
with $\tilde{v}_1=\cos(z_1)$. Importantly, the pure velocity of SVDS should be the relative velocity between the soliton center and plane wave background. Since the units are dimensionless of Eq.~\eqref{eq:Hirota}, therefore the velocity of plane wave background equals to the value of background wavenumber, based on the  quantum mechanics theory \cite{Landau}. Thus, the velocity of SVDS is expressed as
\begin{eqnarray}
\label{v1}
v=\beta(4c^2\tilde{v}_1^2-6ac\tilde{v}_1+3a^2+2c^2)-c\tilde{v}_1,
\end{eqnarray}
It should be emphasized that the wavenumber $a$ cannot be ignored by the Galilean transformation in HE.  Meanwhile, owing to the Eq.~\eqref{v1} is a quadratic function of $\tilde{v}_1$, the variations of velocity will be nonlinear. Therefore, the upper and lower speed limits are not only one, as summarized in Table \ref{v1range}. As we can see, the velocity ranges depend on not only the sign of high-order nonlinearity $\beta$ but also the background wavenumber $a$.  With fixing the sign of $\beta$, the velocity ranges can be categorized into four classes according to the regions of wavenumber $a$, which is greatly different from that of SVDS in NLSE (i.e. for $\beta=0, |v|<c$). That is because the background wavenumber can be removed by the Galilean transformation in NLSE, so it doesn't come into the properties of nonlinear modes \cite{nop,Kivshar}. In contrast to the dark soliton, the velocity of bright soliton is inversely proportional to the strength of high-order nonlinearity and its velocity range is totally unrestricted in HE \cite{Radhakrishnan,solution7,HE1}, similar to that of in NLSE \cite{HE1}. For the other nonlinear waves of HE, such as rational solutions and breathers, although the background wavenumber cannot also be neglected by the Galilean transformation, it just affects the velocity value without determining the velocity range \cite{solution2,CLiu}.

Generally, there is a close connection between the velocity and phase jump. For a dark soliton of NLSE, there is a one-to-one correspondence between the velocity and phase jump \cite{zhaods}. The discussions of the velocity properties imply us both the high-order effects and background wavenumber are crucial to the physical properties of a dark soliton in HE. Then we would like to investigate the phase characters of the SVDS.

\subsection{Two relations between the phase jump and velocity}

\begin{figure}[htp]
\centering
\includegraphics[width=85mm]{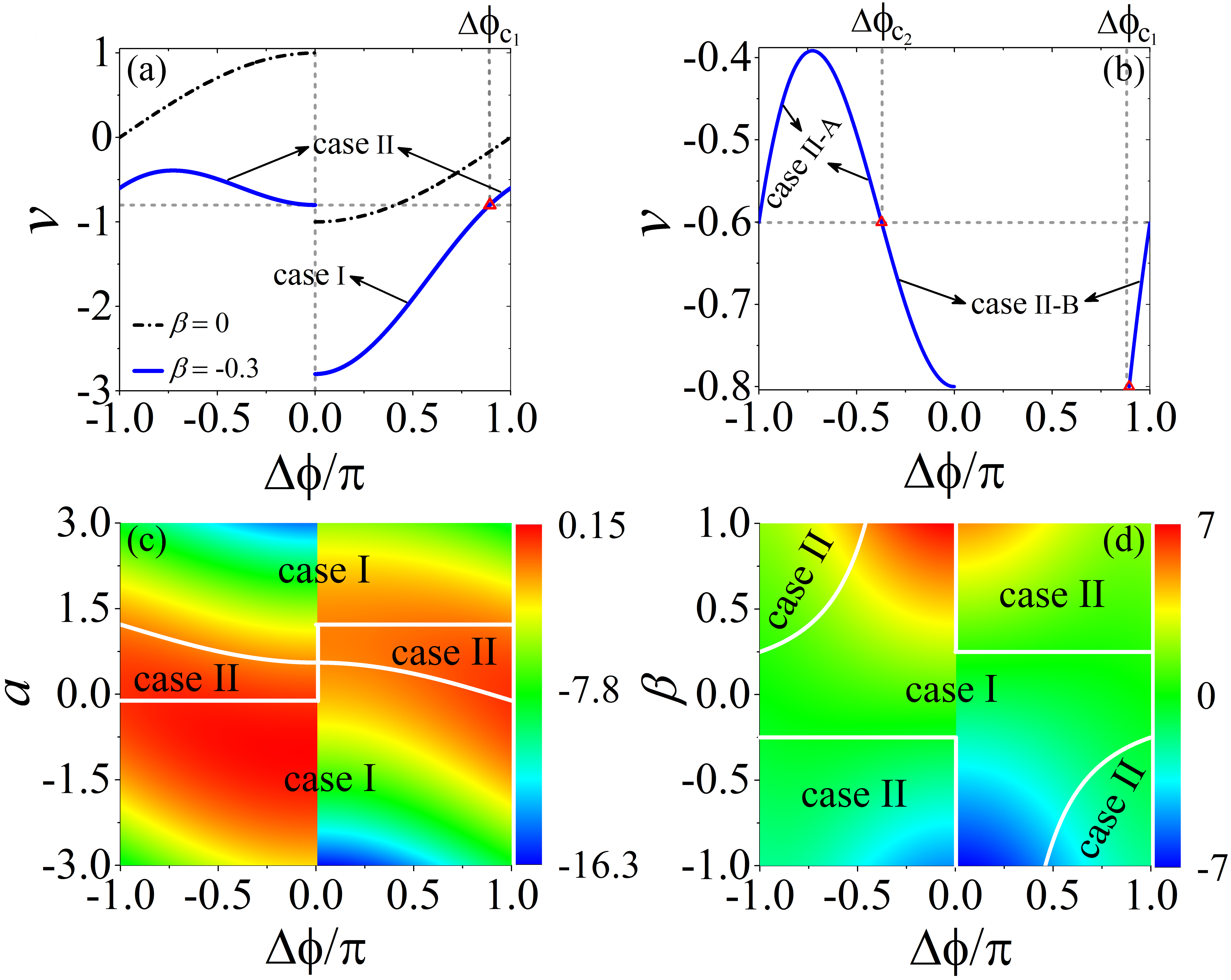}
\caption{Two cases of relations between phase jump ($\Delta\phi$) and velocity ($v$) of an SVDS in HE.  Case I: there is one-to-one match between phase jump and velocity.  Case II: an SVDS with the same velocity admits two different phase jumps. (a) The variations of velocity with phase jump. The blue solid line is obtained with $a=0$ and $\beta=-0.3$ in HE, and the black dashed dotted line corresponds to the NLSE with $\beta=0$. The red symbol corresponds to the critical phase jump between these two cases, $\Delta\phi_{c_1}=2\arccos(1/6)$. (b) Two subcases of the case II shown in Fig.~\ref{Fig1}(a). There exists the second critical phase jump to distinguish the case II-A and the case II-B, $\Delta\phi_{c_2}=-2\arccos(5/6)$. (c) The variations of velocity vs  phase jump $\Delta\phi$ and background wavenumber $a$ with fixing $\beta=-0.3$. (d) The change of velocity vs phase jump $\Delta\phi$ and  high-order nonlinearity $\beta$ with fixing $a=0$. In the phase diagrams Fig.~\ref{Fig1}(c) and (d), the case II is admitted in the regions surrounded by the white solid lines. The outside of these regions is the case I.}\label{Fig1}
\end{figure}

The phase jump of an SVDS is calculated as $\Delta\phi=\int_{-\infty}^{+\infty}\frac{2\tilde{w}^2_1\tilde{v}_1}{1-2\tilde{w}_1^2+\cosh[2\tilde{w}_1x]}\mathrm{d}x
\!=\!2\arctan(\frac{\tilde{w}_1}{\tilde{v}_1})$. Since $z_1\!\in\!(0,\pi\!)$, therefore $\Delta\phi\in\!(-1,1)$.  Then, based on the Eq.~\eqref{v1}, the relations between the phase jump and velocity of the SVDS in HE take the following forms
\begin{subequations}
\begin{gather}
 v= \beta[4+3 a^2+2 \cos(\Delta\phi)]-\cos\Big(\frac{\Delta\phi}{2}\Big)(6 a \beta+1)\label{phi:conditions}\\
 \rm{for}\,\,\,\,\, 0<\Delta\phi<1,\nonumber\\
 v=\beta[4+3 a^2+2 \cos(\Delta\phi)]+\cos\Big(\frac{\Delta\phi}{2}\Big)(6 a \beta+1),\label{phi:conditions2}\\
  \rm{for}\,\,\,\,\, -1<\Delta\phi<0,\nonumber
\end{gather}
\end{subequations}
Obviously, the velocity is co-determined by three physical parameters: the phase jump $\Delta\phi$, the background wavenumber $a$ and the high-order nonlinearity coefficient $\beta$. These two expressions provide a direct way to analyze the phase properties of SVDS in HE. Strkingly, we find that there are two cases of relations between the phase jump and the velocity for the SVDS (as shown in Fig.~\ref{Fig1}):
\begin{itemize}\label{case1}
\setlength{\itemsep}{1pt}
     \item[$\centerdot$]  case I: there is a one-to-one correspondence relation between  phase jump and velocity for an SVDS in HE, which is similar to that in the NLSE.
     \item[$\centerdot$]  case II: there are two different phase jumps for an SVDS with the same velocity in HE, which is not admitted in NLSE.
\end{itemize}
\noindent Next, we will investigate these two cases in detail.

We shall first look briefly at the variations of velocity with phase jump by taking $\beta=-0.3$ and $a=0$, as shown in Fig.~\ref{Fig1}(a) with blue solid line. For comparison, we also show the one-to-one correspondence between phase jump and velocity of an SVDS in NLSE ($\beta=0$) with the black dashed dotted line. It is seen that even if the background wavenumber is not considered, the emergence of the high-order effects also induces the change of velocity with the phase jump to be distinctly different from that of dark solitons in NLSE. Particularly, there exists a critical phase jump $\Delta\phi_{c_1}=2\arccos(1/6)$ labeled by the red triangle symbol, which divides the relations between phase jumps and velocity into the two cases mentioned above.
It is seen that when $\Delta\phi\in [0,\Delta\phi_{c_1})$, the phase jump and velocity of SVDS satisfy the case I relation. Interestingly, when $\Delta\phi\in(-\pi,0)\cap(\Delta\phi_{c_1},\pi)$, an SVDS with the same velocity can admit two distinct phase jumps, namely, it meets the case II relation. This phase property has never been reported in pervious literature.

Moreover, the case II includes two subcases as shown in Fig.~\ref{Fig1}(b), which is distinguished by the second critical phase jump $\Delta\phi_{c_2}=-2\arccos(5/6)$ marked as the left red triangle symbol. For the case II-A, $\Delta\phi\in (-\pi,\Delta\phi_{c_2})$, the two distinct phase jumps have the identical sign but the different values for an SVDS with the same velocity. While for the case II-B, $\Delta\phi\in (\Delta\phi_{c_2},0)\cap (\Delta\phi_{c_1},\pi)$, both the values and the signs of two phase jumps are different. Such abundant phenomena greatly deepen our knowledge of the phase properties for the dark solitons in HE.

More interestingly, we find that background wavenumber $a$ and high-order effects coefficient $\beta$ have a significant impact on the existence regions of the case II. For example, we demonstrate the changes of velocity with phase jump and background wavenumber in Fig.~\ref{Fig1}(c) with fixing $\beta=-0.3$. The existence regions of the case II are surrounded by the white solid lines. It is shown that the case II can emerge when the background wavenumber $a\in(-\frac{1}{9},\frac{11}{9})$ and the phase jump in some ranges. The related critical phase jump $\Delta\phi_c$ will be varied with the background wavenumber. For $a\in(-\frac{1}{9},\frac{5}{9})$, the case II is existed when $\Delta\phi\in(-\pi,0]\cap(\Delta\phi_c,\pi)$ with $\Delta\phi_c=2\arctan\Big[\frac{\sqrt{(5-9a)(9a+7)}}{9a+1}\Big]$. For $a\in[\frac{5}{9},\frac{11}{9})$, the case II can be admitted when $\Delta\phi\in(-\pi,\Delta\phi_c)\cap[0,\pi)$ with $\Delta\phi_c=2\arctan\Big[\frac{\sqrt{(17-9a)(9a-5)}}{9a-11}\Big]$. The outside of these regions, the phase jump and velocity of the SVDS satisfy the case I, which is the same as the one in NLSE.

We further study the influence of the high-order nonlinearity on the existence areas of the case II, with taking the background wavenumber $a\!=\!0$. The result has been shown in Fig.~\ref{Fig1}(d). Considering the high-order effects coefficient is a small value generally, we set $\beta\in[-1,1]$ herein. It is seen that there are two discontinuous regions of $\beta$ for the case II. One is $\beta\in[-1,-\frac{1}{4})$, with $\Delta\phi\in(-\pi,0]\cap(\Delta\phi_c,\pi)$ and $\Delta\phi_c=-2\arctan\Big(\frac{\sqrt{-1-8\beta}}{1+4\beta}\Big)$. The other is $\beta\in(\frac{1}{4},1]$, with  $\Delta\phi\in(-\pi,\Delta\phi_c)\cap[0,\pi)$ and $\Delta\phi_c=2\arctan\Big(\frac{\sqrt{-1+8\beta}}{1-4\beta}\Big)$. Naturally, the parameter spaces except these regions correspond to the case I. It is clear that when the background wavenumber is not considered, there is the case I and only the case I when $|\beta|<\frac{1}{4}$. In this regime, the properties of SVDS are basically the same as the ones in NLSE. The two subcases of case II as mentioned in Fig.~\ref{Fig1}(b) are also existed in Fig.~\ref{Fig1}(c) and (d), we won't elaborate herein.

Above discussions clearly illustrate that an SVDS with the same velocity can admit two distinct phase jumps under certain conditions in HE, in sharp contrast to the dark solitons with only the second-order dispersion and self-phase modulation, which admits a one-to-one match between the velocity and phase jump. Recently, the topological phases of nonlinear waves such as dark solitons \cite{zhaods,Qinds}, bight solitons \cite{Wu}, rogue waves and breathers \cite{zhaorw} have garnered much attention. It was reported that their phase jump are determined solely by the topological vector potentials, which are defined on the complex plane and are periodic along the imaginary axis. It inspires us to investigate the topological vector potentials underlying the distinct phase jumps to further understand the phase characters of the SVDS in the region of the case II, by utilizing the topological vector potential theory proposed in Refs.~\cite{zhaorw,zhaods} directly.

\subsection{The topological vector potentials underlying the phase jumps}\label{IIC}

We first get integrand function $F[x]=\partial_x\varphi$ in the area integral, where $\varphi$ is the phase of wave function $q^{[n]}$ (by expressing the $n$-dark solitons solution Eq.~\eqref{eq:q[n]} as $q^{[n]}=|q^{[n]}|e^{\mathrm{i}\varphi}$ with leaving out the phase of background). Then, we introduce a function $F[z]$, which is obtained by extending the real coordinate variable $x$ to the complex coordinate space $z=x+\mathrm{i}y$ of $F[x]$. The total phase jump can be described by an integral of effective vector potential $\mathbf{A}$ along the real axis (i.e. the $x$ axis). Such an integral corresponds to a circle integral, $\Re\left(\oint_CF[z]\mathrm{d}z\right)=\Re\left(\oint_C(u[x,y]+\mathrm{i}w[x,y])(\mathrm{d}x+\mathrm{i}\mathrm{d}y)\right)=\oint \mathbf{A}\cdot\mathrm{d}\mathbf{r}$, where $\Re$ expresses the real part of this integral, $\mathbf{A}=u[x,y]\mathbf{e}_x-w[x,y]\mathbf{e}_y$ and $\mathrm{d}\mathbf{r}=\mathrm{d}x\mathbf{e}_x+\mathrm{d}y\mathbf{e}_y$. Then, the corresponding magnetic filed is derived as $\mathbf{B}=\nabla\times \mathbf{A}$. It provides a way to explore density zeros ($|q^{[n]}(z_N)|^2=0$) of dark solitons in the extended complex plane and then understand the topological properties underlying their phase jumps. The density zeros $z_N$ are the singularities of $F[z]$ (marked as $z_N=x_N+\mathrm{i} y_N$, $N$ is an integer). Consequently, these singular points constitute monopole fields which appear in pairs. Based on the Cauchy integral formula, the magnetic flux at each of monopoles is calculated as $\Omega=\int_\Gamma f[z]\textrm{d}z=2\pi \mathrm{i} \textrm{\textbf{Res}}{[f(z_N)]}$ \cite{Hayman}, where $\Gamma$ denotes the closed curves encircling the singularity $z_N$. The integral is zero in the complex plane $z$ excepted the singularities. The $\textrm{\textbf{Res}}{[f(z_N)]}=\Omega/2\pi \mathrm{i}$ is the residue. This integral gives rise to a quantized magnetic flux of elementary $\pi$ at each of the monopoles.  The directions of magnetic flux lead to a positive or negative phase jump. Then, we can explore the topological phases of $n$-dark solitons by performing  topological theory on the dark soliton solution $q^{[n]}$ in HE.

\begin{figure}[htp]
\centering
\includegraphics[width=80mm]{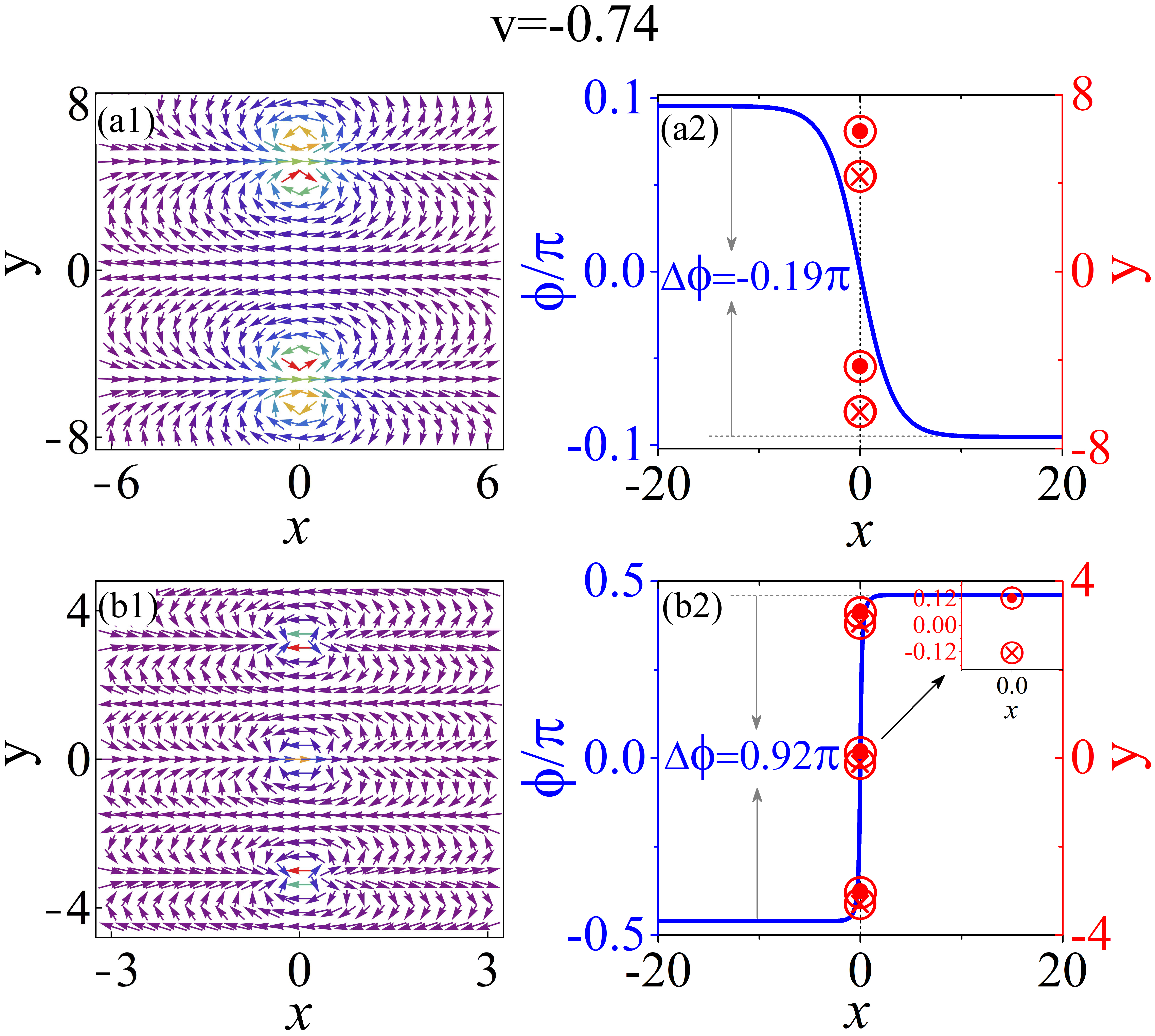}
\caption{Two different topological phases of an SVDS with the same velocity $v=-0.74$ in HE. (a1)-(a2): $\Delta\phi=-0.19\pi$. (b1)-(b2): $\Delta\phi=0.92\pi$. The left panel is the topological vector $\mathbf{A}$, whose corresponding magnetic monopole field $\mathbf{B}$ (red symbols) and phase distribution (blue solid curve) are shown in the right panel.  The paired monopoles are located at $(0,\pm4.3063)$ for (a2) and  $(0,\pm0.1233)$ for (b2) with a period $T=\pi/\sin(z_1)$ along the imaginary $y$. For (a1-a2), $z_1=\arccos[-(25+\sqrt{1045})/60]$. For (b1-b2), $z_1=\arccos[-(25-\sqrt{1045})/60]$. The other parameters are $c=1,a=0,\beta=-0.3,\gamma_1=0$. }\label{Fig2}
\end{figure}

Now we characterize the topology of the SVDS. Based on the solution Eq.~\eqref{onesvds}, the integrand function is expressed as $F[z]=\frac{2\tilde{w}^2_1\tilde{v}_1}{1-2\tilde{w}_1^2+\cosh[2\tilde{w}_1z]}$. Then, we can calculate the vector potential $\mathbf{A}$ by determining the singularities in the complex plane,
\begin{eqnarray}
\label{ZN}
z_N:x_N=0,\,\,\,y_N=\pm y_0+NT  \,\,\,(N=0,\pm1,...)
\end{eqnarray}
where $y_0=\frac{\mathrm{arccosh}(2\tilde{w}_1^2-1)}{2\tilde{w}_1}$ and the period $T=\frac{\pi}{\tilde{w}_1}$. Then, we describe the topological property of an SVDS in the regime of case II. As an example, we exhibit the topological phase of an SVDS with the velocity $v=-0.74$. The related two different topological phases have been presented Fig.~\ref{Fig2}(a1-a2) and (b1-b2), respectively. The vector potentials $\mathbf{A}$ are shown in the left panel, and the corresponding magnetic monopole fields $\mathbf{B}$ (red symbols) and phase distributions $\phi$ (blue solid line) are depicted in the right panel. The pairs of positive and negative magnetic fluxes emerge periodically along the imaginary axis $y$. The monopoles with $+\pi$ and $-\pi$ magnetic fluxes are indicated by $\bigodot$ and $\bigotimes$, respectively. Obviously, these two topological fields in the extending complex plane $z$ are completely different, such as the position, period and even the direction of the magnetic monopoles. Then, the different topological vector potentials give rise to the distinct phase distributions. Therefore, the corresponding phase jump is different. One is $\Delta\phi=-0.19\pi$, the other is $\Delta\phi=0.92\pi$.

Importantly, the relation of case II between the velocity and phase jump shown in Fig.~\ref{Fig1} also indicates that two different SVDSs with the identical velocity are admitted in HE. Namely, two parallel SVDSs can be obtained in the region of case II. Such relation for the SVDS can physically explain the generation mechanism of previously reported two parallel SVDSs in HE \cite{Lou,zhangds,solution6}. When two valleys of these SVDSs are overlapped, the two parallel SVDSs become a DVDS \cite{zhangds}.  One the other hand, the velocity expression Eq.~\eqref{v1} is a quadratic function of $\tilde{v}_1$. Therefore, the two parallel SVDSs can be mathematically constructed with choosing different spectral parameters \cite{Lou,zhangds,solution6}, which is not admitted in NLSE. Based on the abundant phase characters of the SVDS demonstrated above, we can expect that the phase properties of DVDS will be more interesting in HE.

\begin{table}[htp]
\centering
\renewcommand\arraystretch{1.8}
\caption{The velocity ranges of DVDS. The explicit expressions of $v(z_1=z_e)$, $v(z_1=0)$ and $v(z_1=\pi)$ have been given in Table \ref{v1range}. }
\footnotesize
\label{tab}
\begin{tabular}{c|c|c|c}
 \hline
  $\beta$ & $a$ & $z_1$ & $v$ \\
  \hline
  \multirow{2}{*}{$\beta>0$} & $\big[-\frac{1}{6\beta},\frac{8c\beta-1}{6\beta}\big)$ & $\big(0,\arccos(\rho-1)\big)$ & $\big(v(z_1=z_e),v(z_1=\pi)\big)$  \\
  \cline{2-4}
  & $\big(\frac{-8c\beta-1}{6\beta},-\frac{1}{6\beta}\big)$ & $\big(\arccos(\rho+1),\pi\big)$ & $\big(v(z_1=z_e),v(z_1=0)\big)$  \\
  \hline
  \multirow{2}{*}{$\beta<0$} & $\big[-\frac{1}{6\beta},\frac{8c\beta-1}{6\beta}\big)$ & $\big(0,\arccos(\rho-1)\big)$ & $\big(v(z_1=\pi),v(z_1=z_e)\big)$  \\
  \cline{2-4}
  & $\big(\frac{-8c\beta-1}{6\beta},-\frac{1}{6\beta}\big)$ & $\big(\arccos(\rho+1),\pi\big)$ & $\big(v(z_1=0),v(z_1=z_e)\big)$  \\
  \hline
\end{tabular}
\end{table}
\section{The phase characters of a double-valley dark soliton}

\begin{figure}[htp]
\centering
\includegraphics[width=80mm]{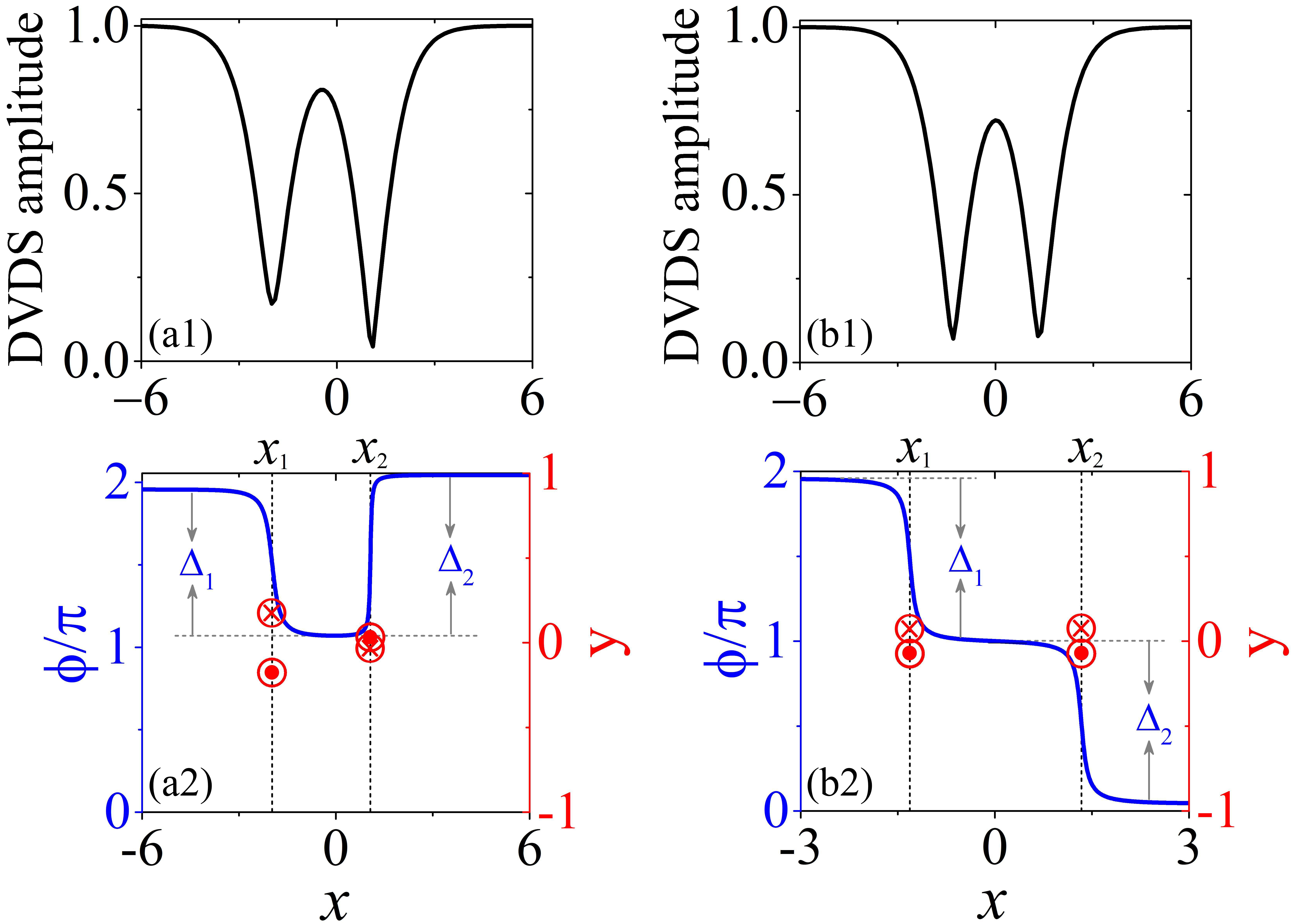}
\caption{Top panel: intensity distributions of a DVDS. (a1) for an asymmetric DVDS ($\gamma_2=-0.1$), (b1) for a symmetric DVDS ($\gamma_2=-1.02$). The other parameters are  $z_1=1.5$, $z_2=1.77931$, $\gamma_1=-1$, $a=0.742485$, $c=1$, $\beta=-0.2$. Bottom panel: phase jumps  (blue solid curve)  and its corresponding magnetic field distribution (red symbols). For (a2), the singularity is located at $(-1.98,\pm0.18)$, $(1.07,\pm0.03)$.   For (b2), the singularities are located at $(-1.31,\pm0.07)$, $(1.34,\pm0.07)$. }\label{Fig3}
\end{figure}

A general DVDS solution in compact form is given by
\begin{equation}
 \label{eq:double}
 \begin{aligned}
q^{[2]}=c\frac{\varrho^2{\rm e}^{-\eta_1-\eta_2}+{\rm e}^{\eta_1+\eta_2}+2\cosh(\eta_1-\eta_2)}{\varrho^2{\rm e}^{-\xi_1-\xi_2}+{\rm e}^{\xi_1+\xi_2}+2\cosh(\xi_1-\xi_2)}{\rm e}^{\mathrm{i}\theta},
 \end{aligned}
 \end{equation}
where $\varrho\!=\!\sin[(z_1-z_2)/2]/\sin[(z_1+z_2)/{2}], \eta_i\!=\!\xi_i+\mathrm{i}z_i, \xi_{i}=\tilde{w}_i(x-v_1t+\gamma_i),\tilde{w}_i=c\sin(z_i), i=1,2$.  $\tilde{w}_i$ is width-dependent parameter. Two free parameters $\gamma_1$ and $\gamma_2$ are introduced to adjust the overlap degree between two valleys of DVDS.  The velocity of the soliton center is obtained by $v_2=v_1$ to meet the conditions of parallel transmission of two SVDSs, based on the Eq.~\eqref{v_j}-Eq.~\eqref{z1:conditions}. Thus, the velocity expression of DVDS is identical to the Eq.~\eqref{v1}. To make a clear understanding of the velocity feature of DVDS, we calculate the velocity ranges of DVDS carefully with satisfying the constraint conditions. The results have been summarized in Table.~\ref{tab}. It is seen that when the sign of the high-order effects is fixed, the velocity ranges can be classified into two categories, which are the same as the final two items for the SVDS shown in Table.~\ref{v1range}. The difference is that there are restrict conditions for the parameters $z_1$ and $z_2$, which determine the spectral parameters. Recently, it was reported that the width-dependent parameters of solitons significantly affect the velocity ranges of DVDS in the Manakov model \cite{Qinds}. On the contrary, the velocity ranges of DVDS of HE depend on the high-order nonlinearity coefficient and background wavenumber. In a certain velocity, the intensity profile of DVDS is asymmetric generally but symmetric for $\gamma_2=\gamma_1+\frac{\tilde{w}_1-\tilde{w}_2}{2\tilde{w}_1\tilde{w}_2}\ln(\rho)$. We have shown an example in Fig.~\ref{Fig3}(a1) and (b1), respectively.

We now extend our analysis on the phase characters to the DVDS. Since the two parallel SVDSs that constitute the DVDS must be in the region of the case II, thus the phase jumps and velocity of these two SVDSs can be in the two subcases of the case II, similar to that shown in Fig.~\ref{Fig1}(b). Therefore, the phase distribution structures of the DVDS will include two types. One is double-step type when two parallel SVDSs in the regime of the case II-A, the other is U-shaped type for the two parallel SVDSs in the region of the case II-B. Then, we will investigate these two types of phase properties of DVDS in detail.

For examples, we study the topological phase properties of the DVDSs shown in Fig.~\ref{Fig3}(a1) and (b1). To facilitate analysis, we make the phase of DVDS at $x\rightarrow-\infty$ remain unchanged, based on the expression $q^{[2]}(x\rightarrow-\infty)=\mathrm{e}^{-\mathrm{i}(z_1+z_2)+\mathrm{i}n\pi}$ ($n$ is an arbitrary integer). Then, we exhibit the corresponding phase distributions in Fig.~\ref{Fig3}(a2) and (b2) with blue solid curves. Obviously, the phase distributions of these DVDSs are distinct. For asymmetric DVDS, its two phase jumps are in opposite directions forming a U-shaped distribution structure, as shown in Fig.~\ref{Fig3}(a2), which has not previously been observed. For the symmetric DVDS, the phase distribution exhibits an apparent double-step structure, as shown in Fig.~\ref{Fig3}(b2), which is similar to that of the DVDS obtained in Manakov systems \cite{zhaods,Qinds}. Based on the topology of dark solitons, the abundant phase characters of DVDS will be also closely related to the underlying topological vector potentials.

Then we further analyze their magnetic monopole fields associated with the vector potential in an extended complex plane, as shown in Fig.~\ref{Fig3}(a2)-(b2) with red symbols. The paired monopoles are located on the imaginary axis with a complicated period. Here, we only show a pair of monopoles with positive and negative flux near the real axis. As it can be seen, two pairs of monopoles are scattered on two separate lines $x=x_1$ and $x=x_2$. The integral of the vector potential corresponding to each line will lead to a phase jump $\Delta_j$ at $x=x_j$ for $j={1,2}$. The total phase jump of a DVDS is given by the sum of two phase jumps.

\begin{figure}[htp]
  \centering
  \includegraphics[width=52mm]{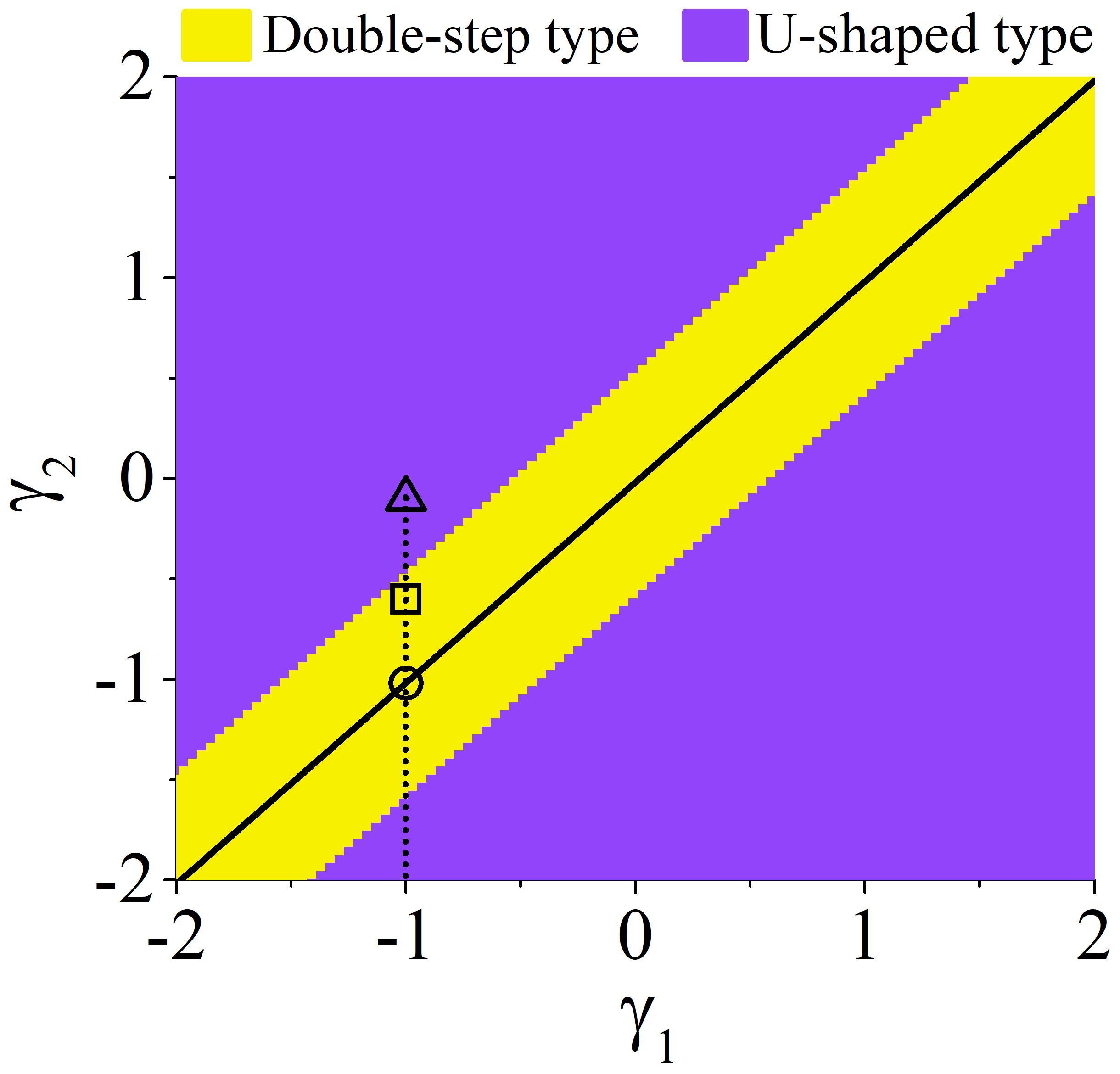}
\caption{A phase diagram for the phase properties of DVDS. The yellow area represents to the double-step type phase distribution, and the purple area stands for the U-shaped type phase distribution. The black solid line correspond to the symmetric DVDS. The triangle and circle symbols correspond to Fig.~\ref{Fig3}(a2) and Fig.~\ref{Fig3}(b2), respectively. The parameters are same as Fig.~\ref{Fig3} (excepting $\gamma_1$ and $\gamma_2$). This picture also describes the phase properties of DVDS before colliding with SVDS, in which the square symbol is related to the DVDS ($``S_2"$ soliton) shown in Fig.~\ref{Fig5}.
}\label{Fig4}
\end{figure}

It is noteworthy that, for the asymmetric DVDS, the direction of paired monopoles on the line $x=x_1$ is opposite to that of on the line $x=x_2$, which gives rise to the two phase jumps in opposite directions, as shown in Fig.~\ref{Fig3}(a2). Moreover, the monopoles located on $x=x_2$ line are colliding with each other, which results in a sharp phase jump $\Delta_2$, closer to $\pi$. We numerically obtain the associated phase steps are $\Delta_1=-0.89\pi$ and $\Delta_2=0.97\pi$ (the coordinates of singularities and the phase jump value are accurate to two decimal places in the present paper). Therefore, the total phase jump of asymmetric DVDS is $\Delta=\Delta_1+\Delta_2=0.08\pi$. These two phase jumps are related to the case II-B mentioned in Fig.~\ref{Fig1}, and their corresponding topologies are similar to those of in Fig.~\ref{Fig2}.

On the contrary, for the symmetric DVDS, two pairs of monopoles distribute in the same direction, as shown in Fig.~\ref{Fig3}(b2). Moreover, each pair of monopoles is preparing to collide on the real axis, which creates a dramatic phase step. Meanwhile, two pairs of monopoles are distributed symmetrically in the complex plane, so the associated two phase steps are the same, $\Delta_1=\Delta_2=-0.96\pi$, which are involved in a special case of the case II-A mentioned in Fig.~\ref{Fig1}. Thus, the total phase jump of the symmetric DVDS is $\Delta=\Delta_1+\Delta_2=-1.92\pi$. These indicate that the different types of phase distributions of DVDS are generated by the distinct topological fields.

Above results have demonstrated that the DVDS with the same velocity can also have two types of phase characters (see caption to the Fig.~\ref{Fig3}), by changing the intensity profiles which can be manipulated by two free parameters $\gamma_1$ and $\gamma_2$. For example, we give a phase diagram for two types of phase characters by changing parameters $\gamma_1$ and $\gamma_2$, and the other parameters are identical to the ones in Fig.~\ref{Fig3}. The results have been presented in Fig.~\ref{Fig4}, which is obtained by judging the sign of the phase gradient flow $F[x]$ numerically. The triangle and circle symbols correspond to the Fig.~\ref{Fig3}(a2) and (b2), respectively. As it can be seen, double-step type phase distribution exists only in a narrow parameter space, among which the black solid line corresponds to the symmetric DVDS. This means the symmetric DVDS always possesses the double-step type phase distribution, while the phase distribution of asymmetric DVDS can be U-shaped type or double-step type. The critical states between these two types are related to the density zero of DVDS in real space. This diagram also indicates that nearly identical intensity profiles of a DVDS with the identical velocity can possess two distinct topologies in adjacent area of the two types. However, the topological phase and the velocity of the DVDS for Manakov system can only be one-to-one correspondence \cite{zhaods,Qinds}.  Our results provide some important complements for understanding the phase properties of dark solitons.

\section{Phase variations of DVDS induced by the collision}

As demonstrated in above section, the phase characters of SVDS and DVDS in single mode fiber with high-order effcts are much more interesting than the ones in the nonlinear system with only second-order dispersion and self-modulation. Our pre-work Ref.~\cite{zhangds} had also shown the unusual inelastic collision dynamics of such DVDSs, by comparing their intensity profiles before and after the collision. Generally, most of studies for solitons collisions are focused on the variations of soliton profiles \cite{Aossey, Huang,Nguyen,Liu,zhaobs,Stalin1,Stalin2,Qinbs,Qindb,lingDT,Qinds,zhangds}, while solitons' phase characters have never been considered. In this section, we will investigate the collision properties of DVDSs by analyzing their topological phases, expecting that phase characteristics can help to identify the collision properties through an alternative way. For simplicity and without loss of generality, we consider the collision between one DVDS and one SVDS, which can be studied based on the solution $q^{[3]}$ (see Eq.~\eqref{eq:q[n]} with setting $n=3$ and $v_2=v_1$). Their collisions can be inelastic or elastic under different conditions. The inelastic collisions make both the intensity profiles and topological phases change of DVDSs, while there is no change for that of the SVDS.

\begin{figure*}[htp]
\centering
\includegraphics[width=172mm]{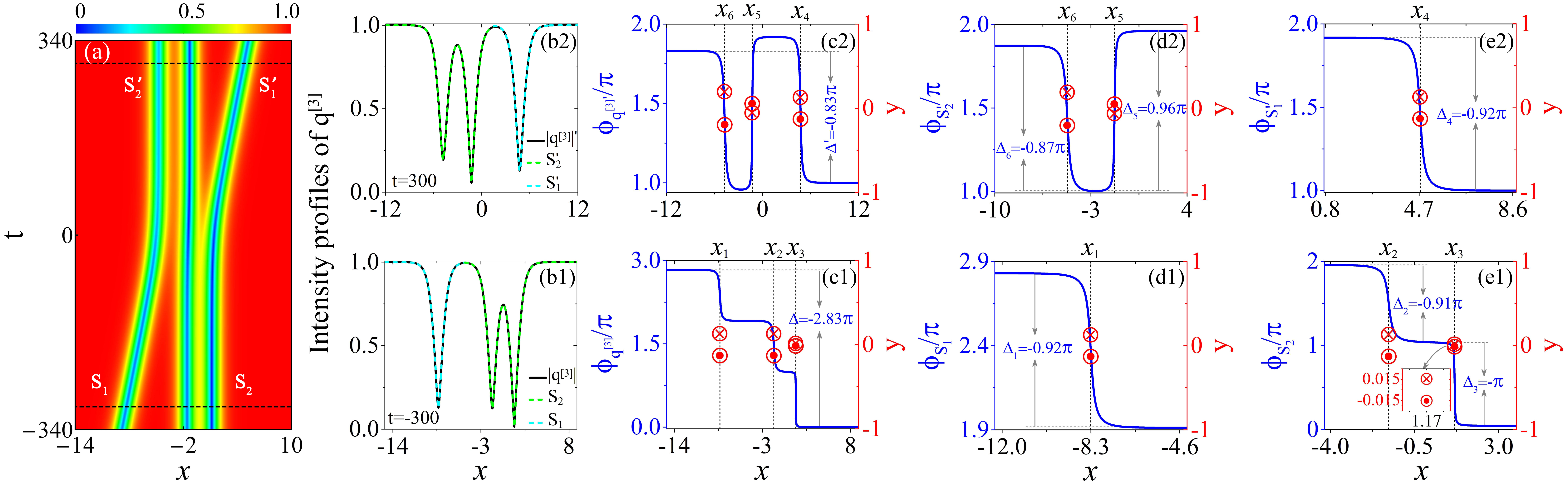}
\caption{Inelastic collision dynamics between an asymmetric DVDS and an SVDS. Before (after) the collision, two solitons, the SVDS and the DVDS are marked as ``$q^{[3]}$", $``S_1"$ and $``S_2"$ (``$q^{[3]'}$", $``S_1'"$ and $``S_2'"$), respectively. (a): temporal-spatial intensity distribution. (b1)-(b2): intensity profiles before $(t=-300)$ and after $(t=300)$ collision. The black solid line, green dashed line and cyan dashed line are plotted by the solution $q^{[3]}$, the asymptotic expression Eq.~\eqref{qs1} of SVDS and Eq.~\eqref{qs2} of DVDS respectively at the fixed $t$. (c1)-(e1): phase distribution and corresponding magnetic field distribution of $q^{[3]}$, $``S_1"$ and $``S_2"$, respectively. The singularities on the line $x=x_1$, $x=x_2$ and $x=x_3$ are located at $(-8.31,\pm0.13)$,
 $(-1.57,\pm0.13)$, $(1.17,\pm0.02)$.
(c2)-(e2): phase distributions and corresponding magnetic field distributions of $q^{[3]'}$, $``S_2'"$ and $``S_1'"$, respectively. The singularities on the line $x=x_6$, $x=x_5$ and $x=x_4$ are located at $(-4.72,\pm0.20)$, $(-1.28,\pm0.05)$, $(4.74,\pm0.13)$. It is seen that both the phase characteristics and intensity profiles of DVDS change remarkably after colliding with SVSD. However, nothing has changed on the latter. The parameters are $c=1,\beta=-0.2,a=0.742485,d=1,z_1=1.5,z_2=1.77931,z_3=1.7,\gamma_1=-1,
\gamma_2=-0.6,\gamma_3=-1$.}\label{Fig5}
\end{figure*}

To get an accurate analysis, firstly we need to analyze the exact solution $q^{[3]}$ through developing the asymptotic analysis technique with assuming $v_2=v_1<v_3$. The detailed asymptotic analyses for $n$-dark solitons involving $n_1$ SVDSs and $n_2$ DVDSs ($n=n_1+2n_2$) have been presented in Appendix \ref{App:prop1} systematically. Here we consider the case with $n_1=1$ and $n_2=1$. Based on the Eq.~\eqref{qk1n} and Eq.~\eqref{qk1p}, the asymptotic expressions of SVDS solution $q_{s_1}$ before and after the collision are given in the following form:
\begin{equation}
\label{qs1}
q_{s_1}^\pm=cL_{s_1}^\pm\left[1-B_3+B_3\tanh(Y_{s_1}^\pm)\right]{\mathrm{e}}^{\mathrm{i}\theta},
\end{equation}
where
\begin{equation*}
\begin{aligned}
&L_{s_1}^-={e}^{-2\mathrm{i}(z_1+z_2)},\,L_{s_1}^+=1,\,K_1^-=\omega_{[3,1]}\omega_{[3,2]},\,K^+=1,\\
&Y_{s_1}^\pm=c\sin{z_3}(x\!-\!v_3t+\gamma_3)\!-\!\ln|K_1^\pm|,\,B_3 =\mathrm{i}\sin{z_3}{\mathrm{e}}^{-\mathrm{i}z_3}.
\end{aligned}
\end{equation*}
The superscript ``$^{\pm}$" in the top right-hand corner represents the asymptotic form of soliton after ($t\rightarrow+\infty$) and before ($t\rightarrow-\infty$) colliding with SVDS. From the asymptotic analysis results Eq.~\eqref{qk2n} and Eq.~\eqref{qk2p}, the asymptotic behaviours of DVDS can be described by the solution $q_{s_2}$, which is given by
\begin{equation}
\label{qs2}
q_{s_2}^\pm=cL_{s_2}^\pm\frac{N_{s_2}^\pm}{D_{s_2}^\pm}\textrm{e}^{\textrm{i}\theta},
\end{equation}
where
\begin{equation*}
\begin{aligned}
N_{s_2}^\pm=&K_2^\pm {\mathrm{e}}^{-\eta_1-\eta_{2}}
+K_3^\pm{\mathrm{e}}^{\eta_{1}-\eta_2}
+K_4^\pm{\mathrm{e}}^{\eta_{2}-\eta_{1}}+{\mathrm{e}}^{\eta_1+\eta_{2}},  \\
D_{s_2}^\pm=&K_2^\pm{\mathrm{e}}^{-\xi_1-\xi_{2}}
+K_3^\pm{\mathrm{e}}^{\xi_{1}-\xi_2}
+K_4^\pm{\mathrm{e}}^{\xi_{2}-\xi_{1}}+{\mathrm{e}}^{\xi_1+\xi_{2}},
\end{aligned}
\end{equation*}
and
\begin{equation*}
\begin{aligned}
&K_2^-=\omega^2_{[1,2]},\,\,\,K_2^+=\omega^2_{[1,2]}\omega^2_{[1,3]}\omega^2_{[2,3]}, \\
&K_3^-=1,\,\,\,K_3^+=\omega^2_{[2,3]},\,\,\,K_4^-=1,\,\,\,K_4^+=\omega^2_{[1,3]},\\
&L_{s_1}^-=1,\,\,\,\,L_{s_1}^+={\mathrm{e}}^{-2\mathrm{i}z_3},\,\,\,\,\omega_{[j,i]}=\frac{\sin[\frac{(z_j-z_i)}{2}]}{\sin[\frac{(z_j+z_i)}{2}]}.
\end{aligned}
\end{equation*}
Above explicit asymptotic expressions provide the most crucial conditions to analyze collision properties exactly and comprehensively.

\subsection{Inelastic collision}

We start with the inelastic interaction between one DVDS and one SVDS. For an example, we show their spatial-temporal intensity distribution in Fig.~\ref{Fig5}(a) with $v_1=0$ and $v_3=0.0125$ (see caption for detailed parameters setting). Before (after) the collision, two solitons, the SVDS and DVDS are marked as ``$q^{[3]}$" (``$q^{[3]'}$"), ``$S_1$" (``$S_1'$") and ``$S_2$" (``$S_2'$"), respectively. The black dashed lines on upper and lower the Fig.~\ref{Fig5}(a) are used to represent the finial state ($t=300$) and initial state ($t=-300$) of these two solitons. In the following, we will analyze the intensity profiles and phase distributions of solitons at these two states. Figure.~\ref{Fig5}(b1)-(b2) are the intensity profiles before and after collision, in which the black solid line, cyan dashed line and green dashed line are plotted based on the solution $q^{[3]}$, $q_{s_1}^{\pm}$ and $q_{s_2}^{\pm}$ at fixed $t$. As it can be observed, the depths of two valleys for DVDS have changed dramatically, while SVDS remain unchanged.  In order to analyze and understand the interaction process comprehensively and deeply, the investigations of phase properties of two solitons before and after the collision are also indispensable. Then, we will conduct the complex extending on expressions Eq.~\eqref{eq:q[n]}, Eq.~\eqref{qs1} and Eq.~\eqref{qs2} to analyze their topological phase, based on the topological vector potential theory mentioned in section \ref{IIC}.

Before (after) the collision, the phase characteristics of $q^{[3]}$ ($q^{[3]'}$), $``S_1"$ ($``S_1'"$) and $``S_2"$ ($``S_2'"$) have been presented in Fig.~\ref{Fig5}(c1-e1) (Fig.~\ref{Fig5}(c2-e2)). In these graphs, the phase distributions in the distribution direction $x$ are presented by the blue solid curves, and their corresponding magnetic field distributions in complex space are depicted by the red symbols. We also only show the paired monopoles which are closer to the $x$-axis herein. Interestingly, the phase jumps of two solitons initially presents triple-step structure in one direction, as shown in Fig.~\ref{Fig5}(c1). Surprisingly, when two solitons accomplish the whole collision process, the total phase distribution turns into a more complex structure composed by three phase jumps in different directions, as exhibited in Fig.~\ref{Fig5}(c2). This dramatic change in the phase jump structure is derived from the underlying topological potentials. By analyzing the magnetic fields associated with the vector potential, we find there are three pairs of monopoles for both cases. Strikingly, three pairs of magnetic monopoles are distributed in the same direction before the collision, while the direction of one of them is reversed after the collision, compared Fig.~\ref{Fig5}(c2) with Fig.~\ref{Fig5}(c1). This motivates us to explore what happened to each of the solitons.

We show the phase distributions and the corresponding magnetic field distributions of SVDS before and after collision in Fig.~\ref{Fig5}(d1) and (e2). It is seen that SVDS maintains single-step structure with one pair of monopoles. Before the collision, monopoles are located on the line $x=x_1$, and the coordinate values in the complex plane is $(-8.31,\pm0.13)$, the associated phase step is $\Delta_1=-0.92\pi$. Then, it moves to the line $x=x_4$, and monopoles are located at $(4.74,\pm0.13)$ after the collision. Obviously, there is only a translation in the $x$-axis for monopoles. Consequently, the phase step is $\Delta_4=\Delta_1=-0.92\pi$. Therefore, both the intensity profile and phase feature of SVDS can be held well after interacting with DVDS.

However, the topological phase of DVDS undergoes a remarkable change, as shown in Fig.~\ref{Fig5}(e1) and (d2). In the initial state, the phase distribution of DVDS is double-step type with two pairs of monopoles in the identical direction.  The paired monopoles on the line $x=x_2$ are located at $(-1.57,\pm0.13)$, which contributes to the phase jump $\Delta_2=-0.91\pi$. Interestingly, the monopoles on the line $x=x_3$ are colliding with each other, tending to merge in the real axis $x$, and the corresponding singularities are located at $(1.17,\pm0.02)$. Then, it leads to a phase jump $\Delta_3=-\pi$ approximately. Due to the coordinate values along the imaginary axis $y$ are too small to see them separately, we further plot the insert for them in Fig.~\ref{Fig5}(e1). Then, we can get the total phase jump of DVDS before the collision is $\Delta_2+\Delta_3=-1.91\pi$.

\begin{figure}[htp]
  \centering
  \includegraphics[width=50mm]{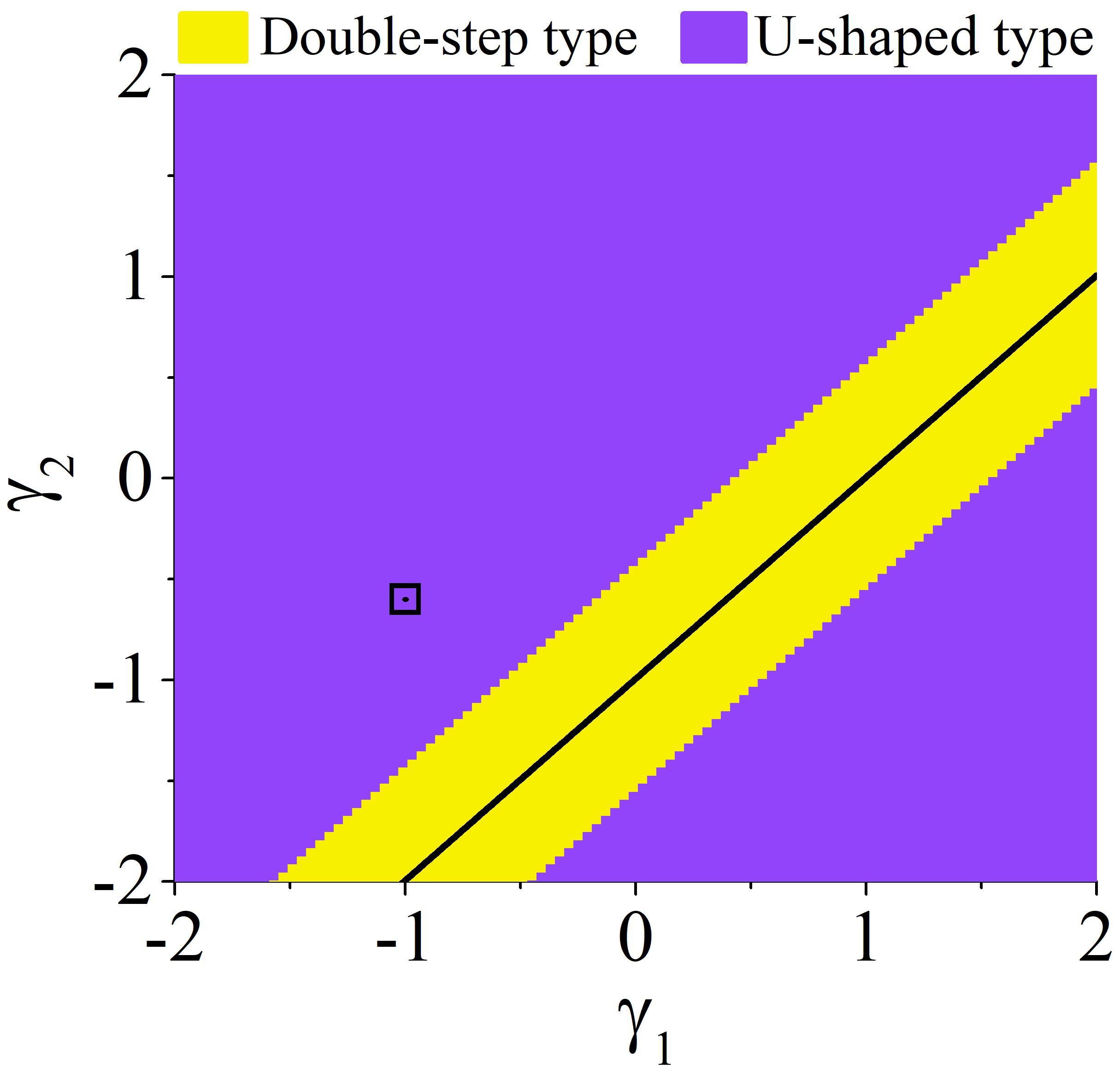}
\caption{A phase diagram for the phase properties of DVDS ($``S_2'"$ soliton) after colliding with SVDS, in which the square symbol represents the phase characteristic of $``S_2'"$ soliton shown in Fig.~\ref{Fig5}. The black line corresponds to the symmetric DVDS. The other parameters are identical with the ones in Fig.~\ref{Fig5}.}\label{Fig6}
\end{figure}

Then, we analyze the phase characteristic of DVDS after the collision, which is shown in Fig.~\ref{Fig5}(d2). Strikingly, the phase distribution of the DVDS is transformed from the double-step type to U-shaped type. Moreover, the collision of paired monopoles is disappeared, in contrast to that of before the collision. Two pairs of monopoles are scattered on two separate lines  $x=x_6$ and $x=x_5$, and their corresponding pointlike magnetic fields are located at $(-4.72,\pm0.20)$ and $(-1.28,\pm0.05)$. These give rise to the two phase steps are $\Delta_6=-0.87\pi$ and $\Delta_5=0.96\pi$ approximately. Thus, the total phase jump of DVDS after the collision is $\Delta_6+\Delta_5=0.09\pi$. This means that the relation of phase jumps and velocity for the two parallel SVDSs constructed the DVDS are transformed from the case II-A to the case II-B before and after the collision. This is an all-important sign of the inelastic collision.

These results imply that inelastic collision of DVDS can cause a striking transformation between two different types of topological phases. Such intriguing phenomenon has never been observed in previous literature. In our pre-work Ref.~\cite{zhangds}, our understanding of the inelastic collision only depends on the variations of intensity profiles, while important phase properties hidden the change of soliton profiles have been ignored. Then we'd like to know whether such phase transition is always existence for the collision dynamics involving the DVDS. For this purpose, we intend to establish phase diagrams for the phase characters of the DVDS before and after the collision, which can be realized by judging the sign of the phase gradient flow $F[x]$ of the asymptotic expressions $q_{s_2}^{\pm}$ Eq.~\eqref{qs2}.

For convenience, we choose the parameters are identical with the ones in Fig.~\ref{Fig5} except parameters $\gamma_1$ and $\gamma_2$. Due to the velocity $(v_1)$ of DVDS center is less than the velocity $(v_3)$ of SVDS center, the asymptotic expressions $q_{s_2}^{-}$ expressed by Eq.~\eqref{qs2} is actually identical to the DVDS solution Eq.~\eqref{eq:double}. It demonstrates that the asymptotic behavior of DVDS before the collision is independent of the SVDS. Meanwhile, owing to the other related parameters $(z_1,z_2, a, c$, and $\beta)$ in Fig.~\ref{Fig5} and Fig.~\ref{Fig3} are the same, thus the phase diagram of phase characters for the DVDS before the collision is identical to the Fig.~\ref{Fig4}. The square symbol in Fig.~\ref{Fig4} corresponds to the $``S_2"$ soliton exhibited in Fig.~\ref{Fig5}(a), which admits the double-step type phase distribution, as shown in  Fig.~\ref{Fig5}(e1). In a similar way, we further give a phase diagram for the DVDS after the collision, with the aid of the asymptotic expressions $q_{s_2}^{+}$ expressed by Eq.~\eqref{qs2}. The results has been presented in Fig.~\ref{Fig6}, where the square symbol corresponds to the $``S_2'"$ soliton depicted in Fig.~\ref{Fig5}(a), which admits the U-shaped type phase distribution, as shown in Fig.~\ref{Fig5}(d2). The black  line is the symmetric DVDS after the collision, and the relation between $\gamma_1$ and $\gamma_2$ is re-calculated as
$\gamma_2=\gamma_1+\frac{(w_1-w_2)\ln(K_{2}^{+})-(w_1+w_2)[\ln(K_4^{+})-\ln(K_3^{+})]}{4w_1w_2}$.

By comparing Fig.~\ref{Fig4} and Fig.~\ref{Fig6}, we can see that the existence region of each type of phase distributions before and after the collision is changed obviously in the $(\gamma_1, \gamma_2)$ plane. During a collision process, the phase variation for the DVDS includes four kinds: double-step type to U-shaped type, U-shaped type to double-step type, U-shaped type to U-shaped type, and double-step type to double-step type. Particularly, the symmetric DVDS always keeps the double-step type phase distribution. These results reveal the extraordinary phase properties behind the inelastic collision and further deepens our understanding of solitons collision in essence. Actually, the emergence of phase transition for the DVDS attribute to the relation of case II shown in Fig.~\ref{Fig1}. It should be mentioned that Fig.~\ref{Fig4} and Fig.~\ref{Fig6} are not the only phase diagrams.  When the related parameters are varied, the corresponding phase diagram will also be changed.

\begin{figure*}[htp]
\centering
\includegraphics[width=172mm]{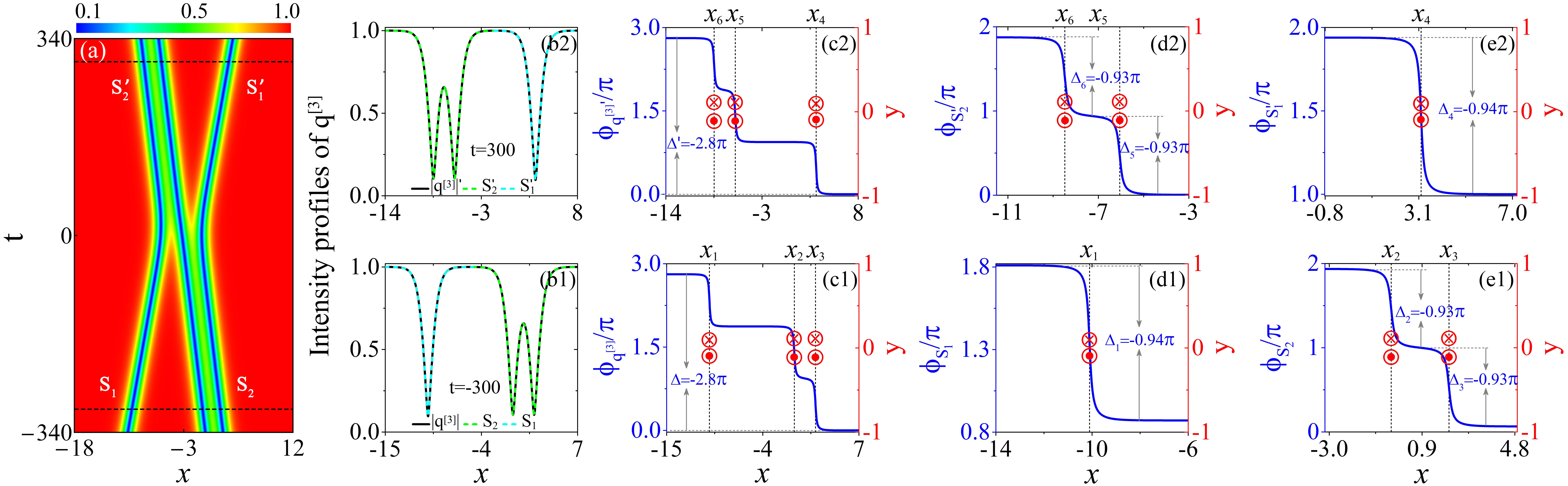}
\caption{Elastic collision dynamics between a symmetric DVDS and an SVDS. Before (after) the collision, two solitons, an SVDS and a DVDS are marked as ``$q^{[3]}$", $``S_1"$ and $``S_2"$ (``$q^{[3]'}$", $``S_1'"$ and $``S_2'"$), respectively. (a): temporal-spatial intensity distribution. (b1)-(b2): intensity profiles before $(t=-300)$ and after $(t=300)$ collision. The black solid line, green dashed line and cyan dashed line are obtained by the two soliton solution Eq.~\eqref{eq:q[n]}, the asymptotic expression Eq.~\eqref{qs1} of SVDS and Eq.~\eqref{qs2} of DVDS ($n=3$), respectively. (c1)-(e1): phase distribution and corresponding magnetic field distributions of $q^{[3]}$, $``S_1"$ and $``S_2"$, respectively. The singularities on the line $x=x_1, x=x_2$ and $x=x_3$ are located at
 $(-10.11,\pm0.10)$, $(-0.40,\pm0.11)$, $(2.03,\pm0.11)$.
(c2)-(e2): phase distribution and corresponding magnetic field distributions of $q^{[3]'}$, $``S_2'"$ and $``S_1'"$, respectively. The singularities on the line $x=x_4, x=x_5$ and $x=x_6$ are located at $(3.20,\pm0.10)$, $(-6.06,\pm0.11)$, $(-8.49,\pm0.11)$. As it can be seen, both the phase properties and intensity profiles of two single valley dark solitons have remained the same after the collision. The parameters are $c=1,\beta=-0.2, a=0.7, d=1, z_1=1.5, z_2=1.84496, z_3=1.66645, \gamma_1=1.1,
\gamma_2=1.06812,\gamma_3=1$.}\label{Fig7}
\end{figure*}

\subsection{Elastic collision}

Above detailed analysis has revealed the intriguing phase transition of DVDS induced by the inelastic collision. It was reported that the collision of DVDS also can be elastic \cite{zhangds}. Then, we will investigate the phase properties of DVDS which is undergone the elastic interaction. The elastic collision condition of DVDS have been proved in the Appendix \ref{symmetric collision}, which is a sufficient condition Eq.~\eqref{elastic condition}. For an example, we show the elastic interaction between a DVDS and an SVDS in Fig.~\ref{Fig7}(a). The speeds of two solitons are $v_3=0.014$ and $v_1=-0.009$.  The meanings of all marks herein are the same as Fig.~\ref{Fig5}, we won't reiterate them here. Based on the expressions $q^{[3]}$, Eq.~\eqref{qs1} and Eq.~\eqref{qs2} and performing the complex extending method, the topological phase of this elastic collision can also be studied well.

The total phase characteristics of two solitons before and after the collision have been shown in Fig.~\ref{Fig7}(c1) and (c2). It is seen that the phase distribution of two solitons keep triple-step structure before and after the collision. By analyzing the magnetic fields associated with the vector potential, we find that there are three pairs of magnetic monopoles in the same direction, scattered on three separate lines at $x_j$ $(j=1,2,3)$.  Considering complexity of vector potential for two solitons solution, we will analyze the phase jump $\Delta_j$ on each line through analyzing the topological phase of each of the solitons.

The phase jump induced by the magnetic field for SVDS before and after collision have been shown in Fig.~\ref{Fig7}(e1) and (d2). A pair of monopoles is initially on the line $x=x_1$, and the corresponding singularity near the real axis is  $(-10.11,\pm0.10)$. Finally, it moves to the line $x=x_4$ located at $(3.20,\pm0.10)$. Therefore, topological property of SVDS has remained the same after colliding with DVDS, associated phase jump $\Delta_4=\Delta_1=-0.94\pi$.

In this case, the DVDS also admits a similar feature, as shown in Fig.~\ref{Fig7}(e2) and (d1). Before the collision, two pairs of monopoles are scattered on the lines $x=x_2$ and $x=x_3$, and the monopoles are located at $(-0.40,\pm0.11)$ and $(2.03,\pm0.11)$. After the collision, they are located at $(-6.06,\pm0.11)$ and $(-8.49,\pm0.11)$ on the line $x=x_6$ and $x=x_5$, respectively. Obviously, two pairs of monopoles only have undergone a translation along the real axis. Moreover, two pairs of monopoles for the DVDS are symmetrically distributed in the complex panel, which leads to a symmetric phase distribution. This also reveals that the associated two parallel SVDSs   have the relation of case II-A mentioned in Fig.~\ref{Fig1}. We get $\Delta_2=\Delta_3=\Delta_6=\Delta_5=-0.93\pi$ numerically, so the total phase jump of DVDS is equal to $-1.86\pi$. Then, we can obtain the total phase jump of two solitons is $\Delta=\Delta'=-0.94\pi-1.86\pi=-2.8\pi$.

These results indicate that we can just analyze the topological phase of each soliton to understand and predict the collision properties, by combining the succinct asymptotic expressions and topological vector potential. In general, the inelastic interactions of solitons bring the variations of both the intensity profiles and topological phase, while the elastic collisions just accompany a translation in the real axis. Similar phenomena can also be observed in the collision between two DVDSs, and even multiple collisions involving the DVDSs, based on the $n$-dark soliton solution Eq.~\eqref{eq:q[n]} and the asymptotic expressions Eq.~\eqref{qk1n}-Eq.~\eqref{qk2p} for $n$-dark solitons including $n_1$ SVDSs and $n_2$ DVDSs ($n = n_1+2n_2$).

\section{Conclusions}

We show that the SVDS with one identical velocity can admit two different phase jumps in HE, which is impossible for dark solitons in NLSE. The two SVDSs satisfied such relations constitute the DVDS reported previously and bring about two types of phase characters for the DVDS. We further uncover the different virtual monopole fields underlying the distinct different phase jumps. Furthermore, we discuss collision properties of DVDSs from topological phase standpoint, which has never been reported before. Particularly, we find that inelastic collision can lead to the transition between two types of phase characters for DVDS. The detailed analyses reveal that the collision properties can be distinguished by analyzing the topological phase, not just intensity profiles. These results provide important supplements and understandings for the phase characters and dynamical properties of dark solitons.

\section*{Acknowledgments}

Y.-H. Qin was supported by the China Postdoctoral Science Foundation (Contract No. 2021M701255), and the National Natural Science Foundation of China (Contract No. 12147170).  L.-C. Zhao was supported by the National Natural Science Foundation of China (Contract No. 12022513, 11775176), and the Major Basic Research Program of Natural Science of Shaanxi Province (Grant No. 2018KJXX-094). Liming
Ling is supported by the National Natural Science Foundation of China (Grant No. 12122105),
the Guangzhou Science and Technology Program of China (Grant No. 201904010362).

\appendix

\section{The derivation of $n$-dark solitons solution}\label{derivation:n-DVDS}

The defocusing HE Eq.~\eqref{eq:Hirota} admits the following Lax pair
\begin{equation}
\label{eq:Lax}
\begin{array}{c}
    \mathbf{\Phi}_x=\mathbf{U}(\lambda;q)\mathbf{\Phi},\,\,\,\,\mathbf{\Phi}_t=\mathbf{V}(\lambda;q)\mathbf{\Phi},
    \end{array}
\end{equation}
where $\lambda$ is real spectral parameter, and
\begin{align*}
\mathbf{U}(\lambda;q)=
     \begin{bmatrix}
      \mathrm{i}\lambda & -\mathrm{i}q^* \\
      \mathrm{i}q & -\mathrm{i}\lambda
    \end{bmatrix},\,\,\,\,
\mathbf{V}(\lambda;q)=
     \begin{bmatrix}
      V_{1,1} & V_{1,2} \\
      V_{2,1} & -V_{1,1}
    \end{bmatrix},
\end{align*}
with
\begin{align*}
& V_{1,1}=\beta(-4\mathrm{i}\lambda^3-2\mathrm{i}|q|^2\lambda+ q_xq^*-q_x^* q)+\mathrm{i}\lambda^2+\frac{\mathrm{i}}{2}|q|^2, \\
 & V_{1,2}=\beta(4\mathrm{i} q^*\lambda^2+2q_x^* \lambda+2\mathrm{i}|q|^2q^*\!-\!\mathrm{i} q_{xx}^*)\!-\!\mathrm{i}q^* \lambda\!-\!\frac{1}{2}q_x^*, \\
 & V_{2,1}=\beta (-4\mathrm{i}q\lambda^2+2q_x\lambda-2\mathrm{i}|q|^2q+\mathrm{i} q_{xx})+\mathrm{i}q\lambda\!-\!\frac{1}{2}q_x.
\end{align*}
To obtain the general solutions of Eq.~\eqref{eq:Hirota}, we start with seed solution $ q^{[0]}=c{\rm e}^{\mathrm{i}\theta}$, with $\theta=ax-bt$ and $b=\beta(a^2+6c^2)a+\frac{1}{2}a^2+c^2$. Then utilizing the transformation $ \mathbf{B}={\rm {\rm diag}}(1,{\rm e}^{-\mathrm{i}\theta})$ to convert the variable coefficient differential equation Eq.~\eqref{eq:Lax} into a constant coefficient equation
\begin{subequations}
\label{eq:lax-2}
\begin{align}
     \tilde{\mathbf{\Phi}}_x&=\mathrm{i}\mathbf{U_1}\tilde{\mathbf{\Phi}}, \,\,\,\,\,\,\, \mathbf{\Phi}=\mathbf{B}^{-1}\tilde{\mathbf{\Phi}},\\
     \tilde{\mathbf{\Phi}}_t&=\mathrm{i}\mathbf{V_1}\tilde{\mathbf{\Phi}}=\mathrm{i}(\varepsilon_1\mathbf{U_1}+\varepsilon_2)\tilde{\mathbf{\Phi}},
\end{align}
\end{subequations}
where
\begin{align*}
\mathbf{U_1}&=\begin{bmatrix}
            \lambda & -c \\
            c & -\lambda-a \\
          \end{bmatrix}, \\
\varepsilon_1&=\beta(-4\lambda^2+2a\lambda-a^2-2c^2)+\lambda-\frac{1}{2}a, \\
\varepsilon_2&=\beta(-2\lambda^2+a_{{1}}\lambda+2c^2)a+\frac{1}{2}a\lambda+\frac{1}{2}c^2,
\end{align*}
The characteristic equation of $ \mathbf{U_1} $ is expressed as
\begin{equation}
      \left(\mu+\frac{a}{2}\right)^2=\left(\lambda+\frac{a}{2}\right)^2-c^2,
      \label{eq:char}
\end{equation}
In order to facilitate the analyses, we introduce the following transformations to parameterized the algebraic equation Eq.~\eqref{eq:char}. Then we have
\begin{align}
\label{mulambda}
\mu+\frac{a}{2}={\rm i} c\sin(z), \,\,\,\, \lambda+\frac{a}{2}=c\cos(z),
\end{align}
with  $0<z<\pi$. By this way, the multivalue problem of the square root can be neatly solved. Thus, the eigenvalue $\mu=-\frac{a}{2}+\textrm{i}\sin(z)$. Based on Eq.~\eqref{eq:lax-2}-Eq.~\eqref{mulambda}, we obtain the vector solution of Eq.~\eqref{eq:Lax} at the spectral parameter $\lambda=\lambda_j=-\frac{a}{2}+c\cos{z_j}, (j=1,\ldots,n)$
\begin{equation*}
\label{eq:Lax sol}
 \mathbf{\Phi}_j=
    \begin{bmatrix}
      {\rm e}^{\mathrm{i}\chi_j} \\
      {\rm e}^{\mathrm{i}(\theta+\chi_j)}(\lambda_j-\mu_j)/c
    \end{bmatrix},
\end{equation*}
with $\chi_j=\mu_jx+(\varepsilon_1\mu_j+\varepsilon_2)t$. Then, the $n$-dark soliton solutions can be derived by applying the $n$-fold DT \cite{lingDT}. By conducting complicated simplification, the explicit expressions of $n$-dark soliton solution can be given in a compact form:
\begin{equation}
\begin{aligned}
\label{eq:q[n]}
q^{[n]}=c\frac{\det(\mathbf{M}_1)}{\det(\mathbf{M})}{\rm e}^{\mathrm{i}\theta},
\end{aligned}
\end{equation}
where
\begin{equation*}
\begin{aligned}
&\mathbf{M}=\left(\frac{{\rm e}^{\mathrm{i}(\chi_j-\bar{\chi}_m)}+\,\delta_{[m,j]}}{{\rm e}^{\mathrm{i}z_j}-{\rm e}^{-\mathrm{i}z_m}}\right)_{1\leq m,j\leq n},\\
&\mathbf{M}_1=\left(\frac{{\rm e}^{\mathrm{i}(\chi_j+z_j-\bar{\chi}_m+z_m)}+\delta_{[m,j]}}{{\rm e}^{\mathrm{i}z_j}-{\rm e}^{-\mathrm{i}z_m}}\right)_{1\leq m,j\leq n},\\
&\chi_j =\mathrm{i}\tilde{w}_j(x-v_j t), \\
&\delta_{[m,j]} =\begin{cases}
      0,& m\neq j \\
       {\rm e}^{2\tilde{w}_m\gamma_m}, & m=j
      \end{cases},\,\,\,\,j=1,2,\cdots,n.
\end{aligned}
\end{equation*}
with $\tilde{w}_j=c\sin(z_j)$, which is the soliton width-dependent parameter. The velocity of soliton center $v_j$ is given by
\begin{equation}
\begin{aligned}\label{v_j}
  v_j=&\beta\left[4 \tilde{v}_j^2-6a\tilde{v}_j+3a^2+2c^2\right]-\tilde{v}_j+a,
\end{aligned}
\end{equation}
and $\tilde{v}_j=c\cos(z_j)$, which is determined jointly by all other parameters in HE. It should be mentioned that, the pure velocity of dark soliton should be the relative velocity between the soliton center and plane wave background. Since the units are dimensionless of Eq.~\eqref{eq:Hirota}, therefore the velocity of plane wave background equals to the value of background wavenumber $a$, based on the quantum mechanics theory. Thus, the velocity of dark soliton is expressed as
\begin{equation}
\begin{aligned}\label{v_j2}
  v_{s_j}=&\beta\left[4 \tilde{v}_j^2-6a\tilde{v}_j+3a^2+2c^2\right]-\tilde{v}_j,
\end{aligned}
\end{equation}
The detailed calculations show that the velocity ranges of SVDS are classed into four cases with the fixing the sign of high-order nonlinearity coefficient $\beta$. The results have been summarized in Table \ref{v1range}.

With choosing $n=1$, Eq.~\eqref{eq:q[n]} is the well-known SVDS solution, which can be simplified as the Eq.~\eqref{onesvds}. It is seen  that the velocity expressions Eq.~\eqref{v_j} and Eq.~\eqref{v_j2} are quadratic function of $\tilde{v}_j$, which means that two SVDSs can admit the identical velocity with choosing two different spectral parameters for $q^{[2]}$, which cannot be realized in NLSE. This provides the possibility to construct the two parallel SVDSs mathematically. When the two valleys of these SVDSs are overlapped, the two parallel SVDSs become a DVDS. In the following, we intend to derive the DVDS solutions.

The first step and the most critical step in constructing DVDS solutions is to make the velocities of two SVDSs are equal, namely, $v_{s_1}=v_{s_2}$ for the solition $q^{[2]}$. To this end, the following constraint condition should be satisfied
\begin{align}
\label{vj:condition}
\cos(z_2)=-\cos(z_1)+\rho,
\end{align}
with $\rho=(6a\beta+1)/(4c\beta)$. Since $ -1<\cos(z_2)<1$, then the following conditions must be satisfied when choosing $z_1$ which determines the spectral parameter $\lambda_1$
\begin{eqnarray}
\label{z1:conditions}
\begin{cases}
     z_1\in (0,\arccos[\rho-1]),& \rho\geq0 \\
     z_1\in (\arccos[\rho+1],\pi) ,&\rho<0
      \end{cases},
\end{eqnarray}
When all parameters meet the constraint conditions Eq.~\eqref{vj:condition} and Eq.~\eqref{z1:conditions}, we can obtain two parallel SVDSs.

The second step to form DVDS is to make two valleys of  two parallel SVDSs be overlapped. In other words, the distance between two valleys cannot be large, otherwise the expression $q^{[2]}$ is still the two SVDSs solution rather than a DVDS solution. This can be realized by adjusting the two free parameters $\gamma_1$ and $\gamma_2$.  Then, we can obtain the DVDS solution based on the two SVDSs solution $q^{[2]}$. The explicit solution expression of DVDS has been simplified as the Eq.~\eqref{eq:double}. The examples for the intensity profiles of DVDS have been shown in Fig.~\ref{Fig3}.

\section{Asymptotic analysis} \label{App:prop1}

We investigate collision behavior of $n$-dark solitons which involves $n_1$ SVDSs and $n_2$ DVDSs ($n=n_1+2n_2$) systematically by developing the asymptotic analysis technique. The $n$-dark solitons are ranked by velocities, supposing $v_1\leq\cdots\leq v_k\leq\cdots\leq v_n$ $(k=1,\cdot\cdot\cdot,n)$. Then we will derive the asymptotic expressions $q_{s_i}$ for the $i$-th soliton $(i=1,\cdot\cdot\cdot,n_1+n_2)$, which can be a SVDS or a DVDS.

\textbf{Case 1}: When $q_{s_i}$ is a SVDS, which is related to the spectral parameter $\lambda_k=-\frac{a}{2}+c\cos(z_k)$. Its propagation direction is controlled by the function $x-v_{k}t=\textrm{const}$, which is contained in the exponential terms ${\rm e}^{\mathrm{i}(\chi_j-\bar{\chi}_j)}$ in the $n$-dark soliton solution Eq.~\eqref{eq:q[n]}. Before the collision ($t\rightarrow -\infty$), there are ${\rm e}^{\mathrm{i}(\chi_j-\bar{\chi}_j)}\rightarrow +\infty$ with $1\leq j<k$ and ${\rm e}^{\mathrm{i}(\chi_j-\bar{\chi}_j)}\rightarrow 0$ with $k<j\leq n$. To get the asymptotic expressions of $q_{s_i}$ before the collision, we need to further eliminate the terms in which ${\rm e}^{\mathrm{i}(\chi_j-\bar{\chi}_j)}\rightarrow\infty$. To this end, ${\rm e}^{\mathrm{i}(\chi_j-\bar{\chi}_j)}$ divide the $j$-th rows of matrices $\mathbf{M_1}$ and $\mathbf{M}$ $ (j=1,\cdots,k-1)$ when $ t\rightarrow -\infty$. Then, the solution Eq.~\eqref{eq:q[n]} can be rewritten as
\begin{eqnarray}
q^{[n]}=c\frac{\det(\mathbf{\hat{M}_1})}{\det(\mathbf{\hat{M}})}\rm{e}^{i\theta},
\end{eqnarray}
where
\begin{equation*}
\begin{aligned}
&\mathbf{\hat{M}}=[\mathbf{\hat{M}}^{(1)},\,\mathbf{\hat{M}}^{(2)},\,\mathbf{\hat{M}}^{(3)}]^\intercal,
\end{aligned}
\end{equation*}
with
\begin{equation*}
\begin{aligned}
&\mathbf{\hat{M}}^{(1)}=\left(\frac{{\rm e}^{\mathrm{i}(\chi_{j}-\chi_i)}}{{\rm e}^{\mathrm{i}z_j}-{\rm e}^{-\mathrm{i}z_i}}\right)_{1\leq i\leq k-1,1\leq j\leq n}, \\
&\mathbf{\hat{M}}^{(2)}=\left(\frac{{\rm e}^{\mathrm{i}(\chi_{j}-\bar{\chi}_{k})}+\delta_{[k,j]}}{{\rm e}^{\mathrm{i}z_j}-{\rm e}^{-\mathrm{i}z_k}} \right)_{1\leq j\leq n},\\
&\mathbf{\hat{M}}=\left(\hat{m}^{(i,j)} \right)_{k+1\leq i\leq n,1\leq j\leq n},\\
&\hat{m}^{(i,j)}=\begin{cases}
      \frac{\delta_i}{{\rm e}^{\mathrm{i}z_i}-{\rm e}^{-\mathrm{i}z_i}} ,& i=j \\
      \frac{{\rm e}^{\mathrm{i}(\chi_{j}-\bar{\chi}_i)}}{{\rm e}^{\mathrm{i}z_j}-{\rm e}^{-\mathrm{i}z_i}} ,& i\neq j
      \end{cases}.
\end{aligned}
\end{equation*}
Here, we define $\delta_{[i,i]}$ as $\delta_i$.   Similarly,
\begin{equation*}
\begin{aligned}
&\mathbf{\hat{M}_1}=\big[\mathbf{\hat{M}_1}^{(1)},\,\mathbf{\hat{M}_1}^{(2)},\,\mathbf{\hat{M}_1}^{(3)}\big]^\intercal,
\end{aligned}
\end{equation*}
where
\begin{equation*}
\begin{aligned}
&\mathbf{\hat{M}_1}^{(1)}=\left(\frac{{\rm e}^{\mathrm{i}(X_{j}-z_j-X_i-z_i)}}{{\rm e}^{\mathrm{i}z_j}-{\rm e}^{-\mathrm{i}z_i}}\right)_{1\leq i\leq k-1,1\leq j\leq n},\\
&\mathbf{\hat{M}_1}^{(2)}=\left(\frac{{\rm e}^{\mathrm{i}(X_{j}-z_j-\bar{X_k}-z_k)}+\delta_{[k,j]}}{{\rm e}^{\mathrm{i}z_j}-{\rm e}^{-\mathrm{i}z_i}}\right)_{1\leq j\leq n},\\
&\mathbf{\hat{M}_1}^{(3)}=\left(\hat{m}_1^{(i,j)} \right)_{k+1\leq i\leq n,1\leq j\leq n},\\
&\hat{m}_1^{(i,j)}=\begin{cases}
      \frac{\delta_i}{{\rm e}^{\mathrm{i}z_i}-{\rm e}^{-\mathrm{i}z_i}} ,& i=j \\
      \frac{{\rm e}^{\mathrm{i}(X_{j}-z_j-\bar{X}_i-z_i)}}{{\rm e}^{\mathrm{i}z_j}-{\rm e}^{-\mathrm{i}z_i}} ,& i\neq j
      \end{cases}.
\end{aligned}
\end{equation*}
By direct calculation, we have
\begin{equation*}
    \begin{split}
    \det(\mathbf{\hat{M}})=&\prod_{i=k+1}^{n}\frac{\delta_{i}}{{\rm e}^{\mathrm{i}z_i}-{\rm e}^{-\mathrm{i}z_i}}\left[{\rm e}^{\mathrm{i}(\chi_{k}-\bar{\chi}_{k})}\det(\mathbf{F}_{k})\right.\\
    &\left.+\frac{\delta_{k}}{{\rm e}^{\mathrm{i}z_k}-{\rm e}^{-\mathrm{i}z_k}}\det(\mathbf{F}_{k-1})\right],
  \end{split}
\end{equation*}
where
\begin{align*}
\mathbf{F}_{l}=\left(\frac{1}{{\rm e}^{\mathrm{i}z_j}-{\rm e}^{-\mathrm{i}z_i}} \right)_{l\leq i,j\leq l},\,\,l=k-1,k.
\end{align*}
Obviously, $ \mathbf{F}_{l} $ is a Cauchy matrix, thus
\begin{align*}
\det(\mathbf{F}_l)=\frac{\prod_{j=2}^{l}\prod_{i=1}^{j-1}({\rm e}^{\mathrm{i}z_j}-{\rm e}^{\mathrm{i}z_i})({\rm e}^{-\mathrm{i}z_i}-{\rm e}^{-\mathrm{i}z_j})}{\prod_{j=1}^{l}\prod_{i=1}^{l}({\rm e}^{\mathrm{i}z_i}-{\rm e}^{-\mathrm{i}z_j})}.
\end{align*}
Similarly,
\begin{equation*}
    \begin{split}
    \det(\mathbf{\hat{M}}_1)=&\prod_{i=k+1}^{n}\frac{\delta_{i}}{{\rm e}^{\mathrm{i}z_i}-{\rm e}^{-\mathrm{i}z_i}}\left[{\rm e}^{\mathrm{i}(\chi_{k}-\bar{\chi}_{k})}\det(\mathbf{G}_{k})\right.   \\
    &\left.+\frac{\delta_{k}}{{\rm e}^{\mathrm{i}z_k}-{\rm e}^{-\mathrm{i}z_k}}\det(\mathbf{G}_{k-1})\right],
  \end{split}
  \end{equation*}
with  $\det(\mathbf{G}_{l})=\prod_{i=1}^{l}{\rm e}^{-2\mathrm{i}z_i}\det(\mathbf{F}_{l})$.
Thus, when $t\rightarrow -\infty$ the asymptotic behavior of SVDS can be derived as
\begin{equation}\label{qk1n}
 q_{s_i}^-=cL_{s_i}^-\left[1-B_k+B_k\tanh(Y_{s_i}^-)\right]{\rm e}^{\mathrm{i}\theta},
\end{equation}
where
\begin{equation*}
\begin{split}
B_k&=\mathrm{i}\sin(z_k)\,\mathrm{e}^{-\mathrm{i}z_k},\\
L_{s_i}^-&=\prod_{m=1}^{k-1}\,\mathrm{e}^{-2\mathrm{i}z_m},\,K_{1}^-=\prod_{i=1}^{k-1}\omega_{[k,i]},\\
Y_{s_i}^-&=c\sin(z_k)(x-v_k t+\gamma_k)-\ln|K_{1}^-|.
\end{split}
\end{equation*}
When $ t\rightarrow +\infty$, we can also derive the asymptotic solution of SVDS in a similar way,
\begin{equation}\label{qk1p}
 q_{s_i}^+=cL_{s_i}^+\left[1-B_k+B_k\tanh(Y_{s_i}^+)\right]{\rm e}^{\mathrm{i}\theta},
\end{equation}
where
\begin{equation*}
\begin{split}
L_{s_i}^+&=\prod_{m=k+1}^{n}\,{e}^{-2\mathrm{i}z_m},\,K_{1}^+=\prod_{i=k+1}^{n}\omega_{[k,i]},\\
Y_{s_i}^+&=c\sin(z_k)(x-v_k t+\gamma_k)-\ln|K_{1}^+|,
\end{split}
\end{equation*}
and $\omega_{[k,i]}=\frac{\sin\left(\frac{z_k-z_i}{2}\right)}{\sin\left(\frac{z_k+z_i}{2}\right)}$.

\textbf{Case 2}: When $q_{s_i}$ is a DVDS, which is related to the spectral parameters $\lambda_k=-\frac{a}{2}+c\cos(z_k)$ and $\lambda_{k+1}=-\frac{a}{2}+c\cos(z_{k+1})$. Then we analyze the asymptotic behavior of DVDS before and after the collision along the propagating direction $ x-v_{k}t=x-v_{k+1}t=\textrm{const}$. Before the collision ($t\rightarrow -\infty$), there are ${\rm e}^{\mathrm{i}(\chi_j-\bar{\chi_j})}\rightarrow +\infty$ for $j<k$,
and ${\rm e}^{\mathrm{i}(\chi_j-\bar{\chi_j})}\rightarrow 0$ for $j>k+1$.
Then, ${\rm e}^{\mathrm{i}(\chi_j-\bar{\chi_j})}$ divide the $j$-th rows of both matrices $\mathbf{M_1}$ and $\mathbf{M}$   $(j=1,\cdots,k-1)$ and taking the limit $t\rightarrow -\infty $, then the solution Eq.~\eqref{eq:q[n]} can be rewritten as
\begin{eqnarray}
q^{[n]}=c\frac{\det(\mathbf{\hat{M}_1})}{\det(\mathbf{\hat{M}})}\textrm{e}^{\mathrm{i}\theta},
\end{eqnarray}
where
\begin{equation*}
\begin{aligned}
&\mathbf{\hat{M}}=[\mathbf{\hat{M}}^{(1)},\, \mathbf{\hat{M}}^{(2)},\,\mathbf{\hat{M}}^{(3)}]^\intercal,
\end{aligned}
\end{equation*}
with
\begin{equation*}
\begin{aligned}
&\mathbf{\hat{M}}^{(1)}=\left(\frac{{\rm e}^{\mathrm{i}(\chi_{j}-\chi_i)}}{{\rm e}^{\mathrm{i}z_j}-{\rm e}^{-\mathrm{i}z_i}}\right)_{1\leq i\leq k-1,1\leq j\leq n}, \\
&\mathbf{\hat{M}}^{(2)}=\left(\frac{{\rm e}^{\mathrm{i}(\chi_{j}-\bar{\chi}_{i})}+\delta_{[i,j]}}{{\rm e}^{\mathrm{i}z_j}-{\rm e}^{-\mathrm{i}z_i}} \right)_{k\leq i\leq k+1,1\leq j\leq n},\\
&\mathbf{\hat{M}}^{(3)}=\left(\hat{m}^{(i,j)} \right)_{k+2\leq i\leq n,1\leq j\leq n},\\
&\hat{m}^{(i,j)}=\begin{cases}
      \frac{\delta_i}{{\rm e}^{\mathrm{i}z_i}-{\rm e}^{-\mathrm{i}z_i}} ,& i=j \\
      \frac{{\rm e}^{\mathrm{i}(\chi_{j}-\bar{\chi}_i)}}{{\rm e}^{\mathrm{i}z_j}-{\rm e}^{-\mathrm{i}z_i}} ,& i\neq j
      \end{cases}
\end{aligned}
\end{equation*}
We can also write $\mathbf{\hat{M}_1}$ as
\begin{equation*}
\begin{aligned}
&\mathbf{\hat{M}_1}=[\mathbf{\hat{M}_1}^{(1)},\,\mathbf{\hat{M}_1}^{(2)},\, \mathbf{\hat{M}_1}^{(3)}]^\intercal,
\end{aligned}
\end{equation*}
with
\begin{equation*}
\begin{aligned}
&\mathbf{\hat{M}_1}^{(1)}=\left(\frac{{\rm e}^{\mathrm{i}(\chi_{j}-z_j-\chi_i-z_i)}}{{\rm e}^{\mathrm{i}z_j}-{\rm e}^{-\mathrm{i}z_i}}\right)_{1\leq i\leq k-1,1\leq j\leq n},\\
&\mathbf{\hat{M}_1}^{(2)}=\left(\frac{{\rm e}^{\mathrm{i}(\chi_{j}-z_j-\bar{\chi_i}-z_i)}+\delta_{[i,j]}}{{\rm e}^{\mathrm{i}z_j}-{\rm e}^{-\mathrm{i}z_i}}\right)_{k\leq i\leq k+1,1\leq j\leq n},\\
&\mathbf{\hat{M}_1}^{(3)}=\left(\hat{m}_1^{(i,j)} \right)_{k+2\leq i\leq n,1\leq j\leq n},\\
&\hat{m}_1^{(i,j)}=\begin{cases}
      \frac{\delta_i}{\mathrm{e}^{\mathrm{i}z_i}-\mathrm{e}^{-\mathrm{i}z_i}} ,& i=j \\
      \frac{{\rm e}^{\mathrm{i}(\chi_{j}-z_j-\bar{\chi}_i-z_i)}}{{\rm e}^{\mathrm{i}z_j}-{\rm e}^{-\mathrm{i}z_i}} ,& i\neq j
      \end{cases}
\end{aligned}
\end{equation*}
Then the determinant of $\mathbf{\hat{M}}$ and $ \mathbf{\hat{M}_1} $ can be written as
\begin{equation*}
    \begin{split}
     &\det(\mathbf{\hat{M}})=\\
    &\prod_{i=k+2}^{n}\frac{\delta_{i}}{{\rm e}^{\mathrm{i}z_i}-{\rm e}^{-\mathrm{i}z_i}}\left[{\rm e}^{\mathrm{i}\left[(\chi_{k}-\bar{\chi}_{k})+(\chi_{k+1}-\bar{\chi}_{k+1})\right]}\det(\mathbf{F}_{k+1}) \right.\\
    &+\frac{\delta_{k}{\rm e}^{\mathrm{i}(\chi_{k+1}-\bar{\chi}_{k+1})}}{{\rm e}^{\mathrm{i}z_k}-{\rm e}^{-\mathrm{i}z_k}}\det(\mathbf{\hat{F}}_{k})+\frac{\delta_{k+1}{\rm e}^{\mathrm{i}(\chi_{k}-\bar{\chi}_{k})}}{{\rm e}^{\mathrm{i}z_{k+1}}-{\rm e}^{-\mathrm{i}z_{k+1}}}\det(\mathbf{F}_{k})\\
&\left.+\frac{\delta_{k}\delta_{k+1}}{({\rm e}^{\mathrm{i}z_k}-{\rm e}^{-\mathrm{i}z_k})({\rm e}^{\mathrm{i}z_{k+1}}-{\rm e}^{-\mathrm{i}z_{k+1}})}\det(\mathbf{F}_{k-1})\right],\\
 &\det(\mathbf{\hat{M}_1})= \\
 &\prod_{i=k+2}^{n}\frac{\delta_{i}}{{\rm e}^{\mathrm{i}z_{i}}-{\rm e}^{-\mathrm{i}z_{i}}}\left[{\rm e}^{\mathrm{i}\left[(\chi_{k}-\bar{\chi}_{k})+(\chi_{k+1}-\bar{\chi}_{k+1})\right]}\det(\mathbf{G}_{k+1}) \right.\\
 &+\frac{\delta_{k}{\rm e}^{\mathrm{i}(\chi_{k+1}-\bar{\chi}_{k+1})}}{{\rm e}^{\mathrm{i}z_{k}}-{\rm e}^{-\mathrm{i}z_{k}}}\det(\mathbf{\hat{G}}_{k})+\frac{\delta_{k+1}{\rm e}^{\mathrm{i}(\chi_{k}-\bar{\chi}_{k})}}{{\rm e}^{\mathrm{i}z_{k+1}}-{\rm e}^{-\mathrm{i}z_{k+1}}}\det(\mathbf{G}_{k})\\
&\left.+\frac{\delta_{k}\delta_{k+1}}{({\rm e}^{\mathrm{i}z_{k}}-{\rm e}^{-\mathrm{i}z_{k}})({\rm e}^{\mathrm{i}z_{k+1}}-{\rm e}^{-\mathrm{i}z_{k+1}})}\det(\mathbf{G}_{k-1})\right],\\
\end{split}
\end{equation*}
where
\begin{equation*}
 \begin{split}
\det(\mathbf{F}_l)&=\frac{\prod_{j=2}^{l}\prod_{i=1}^{j-1}({\rm e}^{\mathrm{i}z_{j}}-{\rm e}^{\mathrm{i}z_{i}})({\rm e}^{-\mathrm{i}z_{i}}-{\rm e}^{-\mathrm{i}z_{j}})}{\prod_{j=1}^{l}\prod_{i=1}^{l}({\rm e}^{\mathrm{i}z_{i}}-{\rm e}^{-\mathrm{i}z_{j}})},\\
\det(\mathbf{\hat{F}}_{k})&=\prod_{i=1}^{k-1}\frac{({\rm e}^{\mathrm{i}z_{k+1}}-{\rm e}^{\mathrm{i}z_{i}})({\rm e}^{-\mathrm{i}z_{i}}-{\rm e}^{-\mathrm{i}z_{k+1}})}{({\rm e}^{\mathrm{i}z_{k+1}}-{\rm e}^{-\mathrm{i}z_{i}})({\rm e}^{\mathrm{i}z_{i}}-{\rm e}^{-\mathrm{i}z_{k+1}})}\\
 &\cdot \frac{1}{{\rm e}^{\mathrm{i}z_{k+1}}-{\rm e}^{-\mathrm{i}z_{k+1}}} \det(\mathbf{F}_{k-1}),\\
\det(\mathbf{G}_{l})&=\prod_{i=1}^{l}{\rm e}^{-2\mathrm{i}z_i}\det(\mathbf{F}_{l}),\,\,\,l=k-1,k,k+1,\\
 \det(\mathbf{\hat{G}}_{k})&=\prod_{i=1}^{k-1}{\rm e}^{-2\mathrm{i}z_i} {\rm e}^{-2\mathrm{i}z_{k+1}}  \det(\mathbf{\hat{F}}_{k}).
\end{split}
\end{equation*}
Thus, the asymptotic behavior of the DVDS at $t\rightarrow -\infty$ is expressed as
\begin{equation}
\begin{split}\label{qk2n}
q_{s_i}^{-}=c\frac{\det(\mathbf{\hat{M}}_1)}{\det(\mathbf{\hat{M}})}\textrm{e}^{\mathrm{i}\theta}=cL_{s_i}^{-}\frac{
N_{s_i}^-}{D_{s_i}^-}\textrm{e}^{\mathrm{i}\theta},
\end{split}
\end{equation}
By performing the similar procedure as presented above, along the trajectory $ x-v_{k}t=x-v_{k+1}t=\textrm{const}$ at $t\rightarrow +\infty$, we get the asymptotic expression of DVDS after the collision as the following form
\begin{eqnarray}\label{qk2p}
q_{s_i}^{+}=c\frac{\det(\mathbf{\hat{M}}_1)}{\det(\mathbf{\hat{M}})}\textrm{e}^{\textrm{i}\theta}=cL_{s_i}^{+}\frac{
N_{s_i}^+}{D_{s_i}^+}\textrm{e}^{\textrm{i}\theta},
\end{eqnarray}
where
\begin{eqnarray*}
N_{s_i}^\pm&\!=\!&
K_2^\pm {\rm e}^{-\eta_{k}-\eta_{k+1}}
+K_3^\pm{\rm e}^{\eta_{k}-\eta_{k+1}}+K_4^\pm{\rm e}^{\eta_{k+1}-\eta_{k}}\\
&&+{\rm e}^{\eta_{k}+\eta_{k+1}},\\
D_{s_i}^\pm&\!=\!&
K_2^\pm {\rm e}^{-\xi_{k}-\xi_{k+1}}
+K_3^\pm{\rm e}^{\xi_{k}-\xi_{k+1}}+K_4^\pm{\rm e}^{\xi_{k+1}-\xi_{k}}\\
&&+{\rm e}^{\xi_{k}+\xi_{k+1}},
\end{eqnarray*}
and
\begin{equation*}
\begin{split}
L_{s_i}^-&=\prod_{m=1}^{k-1}\mathrm{e}^{-2\mathrm{i}z_m},\,\,\,\,\,\,\, \,\, \,\, L_{s_i}^+=\prod_{m=k+2}^{n}\mathrm{e}^{-2\mathrm{i}z_m},\\
K_2^-&=\prod_{j=k}^{k+1}\prod_{i=1}^{j-1}\omega^2_{[j,i]}, \,\,\,\,\,\,\,\, K_2^+=\prod_{j=k}^{k+1}\prod_{i=j+1}^{n}\omega^2_{[j,i]},\\
K^-_3&=\prod_{i=1}^{k-1}\omega^2_{[k+1,i]}, \,\,\,\,\,\,\,\, \,\, \, K^+_3=\prod_{i=k+2}^{n}\omega^2_{[k+1,i]},\\
K^-_4&=\prod_{i=1}^{k-1}\omega^2_{[k,i]}, \,\,\,\,\,\,\, \,\, \,\,  \,\, \,\, \,\,K^+_4=\prod_{i=k+2}^{n}\omega^2_{[k,i]},
\end{split}
\end{equation*}
In addition, we let $ \prod_{m=1}^{k-1}f=1 $ when $k=1 $,
and $ \prod_{m=k+1}^{n}f=1$ when $ k=n $, where $ f=\omega^2$ or $\mathrm{e}^{-2\textrm{i}z_m}$ in above related expressions.

\section{The elastic collision for SVDS and DVDS}\label{symmetric collision}

The asymptotic results Eq.~\eqref{qk1n}, Eq.~\eqref{qk1p}, Eq.~\eqref{qk2n} and Eq.~\eqref{qk2p} provide the fundamental conditions to analyze their collision properties directly and exactly. From the Eq.~\eqref{qk1n} and Eq.~\eqref{qk1p}, we can easily get
\begin{equation*}
\left|q_{s_i}^-\left(x+x_1\right) \right|^2=\left|q_{s_i}^+\left(x+x_2\right) \right|^2,
\end{equation*}
with
\begin{eqnarray*}
x_1=\frac{\ln|K_1^-|}{c\sin(z_k)}, \,\,\,\,\, x_2=\frac{\ln|K_1^+|}{c\sin(z_k)},
\end{eqnarray*}
This demonstrates that SVDS survives its shape after the collision with a phase shift, similar to the solitons' collision reported before. However, DVDS does not admit such property in general case, as shown in Fig.~\ref{Fig5}. Then we would like to know whether the collision of DVDS can also be elastic. We rewrite the Eq.~\eqref{qk2n} and Eq.~\eqref{qk2p} as following form to facilitate our analyses
\begin{equation}
\label{mqk2}
\left|q_{s_i}^\pm\right|^2=c^2\left[1+\partial_{xx}\ln(D_{s_i}^\pm)\right],\\
\end{equation}
When the collision of $ n$-dark solitons is elastic for the DVDS $q_{s_i}$ which is related to the spectral parameters $\lambda_k$  and $\lambda_{k+1}$,
it will only result in a phase shift. Thus, we can assume $\left|q_{s_i}^-(x+c_1)\right|^2=\left|q_{s_i}^+(x)\right|^2,$ where $ c_1 $ is a constant. According to Eq.~\eqref{mqk2}, we just need to let $D_{s_i}^-(x+c_1)=D_{s_i}^+(x)$. Then, we have
\begin{equation}
\begin{split}\label{elastic}
\ln(K_3^+)&=\ln(K_3^-)-c\sin(z_{k+1})c_1,\\
 \ln(K_4^+)&=\ln(K_4^-)-c\sin(z_k)c_1,
\end{split}
\end{equation}
Based on Eq.~\eqref{elastic}, we get
\begin{equation}
\label{elastic condition}
\begin{split}
\frac{\ln(K_3^+)-\ln(K_3^-)}{\ln(K_4^+)-\ln(K_4^-)}=\frac{\sin(z_{k+1})}{\sin(z_k)}.
\end{split}
\end{equation}
Therefore,  under this condition,  the collisions contained DVDSs can be elastic. In other words, this is the sufficient condition, Eq.~\eqref{elastic condition}, of elastic interaction for DVDSs.

\end{document}